\documentclass[twocolumn,amsmath,amssymb]{revtex4}
\usepackage{amsfonts}
\usepackage[usenames]{color}

\newcommand{\bea}{\begin{eqnarray}}
\newcommand{\eea}{\end{eqnarray}}

\begin{document}
\title{Massive-photon electrodynamics and MHD in curved spacetime and cosmology}
\author{Jai-chan Hwang${}^{1}$ and Hyerim Noh${}^{2}$}
\address{${}^{1}$Particle Theory  and Cosmology Group,
         Center for Theoretical Physics of the Universe,
         Institute for Basic Science (IBS), Daejeon, 34051, Republic of Korea
         \\
         ${}^{2}$Theoretical Astrophysics Group, Korea Astronomy and Space Science Institute, Daejeon, Republic of Korea
         }

\begin{abstract}

We study a massive-photon electrodynamics and magnetohydrodynamics (MHD) in the curved spacetime of Einstein's gravity. We consider a Proca-type photon mass and present equations in terms of electric and magnetic (EM) fields and the vector potential. We present the electrodynamics and MHD in the covariant and ADM formulations valid in general spacetime and in linearly perturbed cosmological spacetime. We present wave equations assuming the metric variations are negligible compared with the field variations. Equations are derived without fixing the temporal gauge condition and the gauge transformation properties of the EM fields and the vector potential are presented. Using the post-Newtonian approximation we show the dark Proca field behaves as dust in the non-relativistic limit under the Klein transformation.

\end{abstract}

\maketitle

\tableofcontents

%
%
%
\section{Introduction}

The mass of photons is a fundamental issue in physics that can be constrained only by experiments and observations \cite{Kobzarev-1968, Goldhaber-1971, Chibisov-1976, Byrne-1977, Tu-2005, Goldhaber-2010, Spavieri-2011}. Effect of non-vanishing photon mass was discussed by de Broglie in his 1925 thesis and earlier in \cite{deBroglie-1922, deBroglie-1923, deBroglie-1925}, and later he mentioned ``if [the mass of photons is] not zero then that would be one of the most important constants in physics" \cite{deBroglie-1934}. The idea was elaborated by his disciple Alexandru Proca in the context of the massive spin-1 vector mesons \cite{Proca}; for more related works by Proca, see references in \cite{Goldhaber-1971, Tu-2005, Goldhaber-2010}; for a historical summary, see \cite{Spallicci-2024}. Although Proca himself considered the photon massless, we will consider the Proca's equation for photons.

If the photon has a finite mass it can have dynamical impact on scales larger than the photon Compton wavelength. The static electric and magnetic (EM) fields influence only finite range, and beyond the Compton wavelength scale the fields decay exponentially. The current limit on the photon mass based on the Solar System measurement is quite stringent in the laboratory standard. Using the reduced Compton wavelength, $\lambdabar \equiv \hbar /(mc)$, we have $\lambdabar > 2 \times 10^{13} {\rm cm} \sim 1.3 {\rm AU}$ \cite{Ryutov-2007, Ryutov-2009, Goldhaber-2010, Bhattacharjee-2023, Spallicci-2024}. From this follows the mass limit
\bea
   m < 2 \times 10^{-51} {\rm g}
       \sim 10^{-18} {\rm eV}/c^2.
\eea
There are tighter limits based on the cosmic scale magnetic fields measurements including galaxies, but these are still speculative \cite{Goldhaber-2010} and controversial \cite{Adelberger-2007}.

The additional longitudinal polarization has a negligible effect on the relativistic degrees of freedom \cite{Bass-1955, Murphy-1975}. The modification of the Planck formula is supressed by photon mass square factors and the consequent bound on the photon mass, from laboratory measurements and cosmology, is weaker than the astrophysical bound quoted above \cite{Torres-Hernandez-1985}.

The above limit, limited to the Solar System scale, allows potential importance of the photon mass in the electromagnetic phenomena in cosmic scales, including the cosmological scale. In the latter scale, relativistic effects like cosmic expansion and perturbed order metric may play role. Here, we present massive-photon electrodynamics and magnetohydrodynamics (MHD) in the general curved spacetime of Einstein's gravity, and especially in cosmological spacetime with general linear metric perturbations.

Our main emphasis is the relativistic aspects of the electrodynamics and MHD valid in general curved spacetime and in the linearly perturbed cosmological spacetime. Potential applications of our formulation concern relativistic regime in astrophysics and cosmology involving magnetic field and photons especially beyond photon Compton wavelength, currently around 1AU scale, where the photon mass can play roles. As magnetic fields are everywhere, this may include many astrophysical systems where gravity is strong . Examples are accretion disc and jet systems around supermassive black holes; the event horizon radius of a billion solar mass supermassive black hole is around 19AU. This is an active field with significant numerical advancement \cite{Thorne-1986, Beskin-2010}. For textbooks on relativistic astrophysics ignoring the photon mass, see \cite{Wilson-Mathews-2003, Baumgarte-Shapiro-2010, Gourgoulhon-2012, Shibata-2015, Baumgarte-Shapiro-2021}. We note that our covariant and ADM formulations are valid in general curved spacetime with the hydrodynamic conservation equations and Einstein equations presented in our previous work \cite{HN-2023-EM}. Consideration of the photon mass in astrophysical situations may lead to constraints on the photon mass.

In cosmology, previous studies on the subject including magnetogenesis are mainly based on the Friedmann background without metric perturbations, for reviews see \cite{Grasso-2001, Widrow-2002, Giovannini-2004, Kulsrud-2008, Kandus-2011, Durrer-2013, Subramanian-2016, Vachaspati-2021}. This is because the homogeneous and isotropic symmetry of the background cosmology does not allow EM fields to the linear order. One exception is the case of massive photon where in the non-relativistic limit we have dark matter like behavior, but its not ordinary photons, see Section \ref{sec:Proca-DM}. Even without the photon mass term, Maxwell's equations in the linearly perturbed cosmology without fixing the gauge derived in this work are new in the literature.

As a massive-photon electrodynamics we consider Maxwell equations with a general interaction potential and in particular a Proca-type photon mass term. We present Maxwell equations in curved spacetime using the EM fields and the vector potential. We present the massive-photon electrodynamics and MHD equations using the covariant and the ADM (Arnowitt-Deser-Misner) formulations valid in general curved spacetime and in the linearly perturbed cosmological spacetime. The photon mass effect can be interpreted as effective charge and current densities, and this allows us to use the general exact formulations presented in \cite{HN-2023-EM}.

As the vector field cannot be accommodated in the spatially homogeneous and isotropic Friedmann background, we ignore the gravity of the EM fields in cosmology. Thus, we ignore Einstein's equation in cosmology, and consider Maxwell's equations in a given cosmological spacetime with linear metric perturbation. One exception is the case where a dark Proca field behaves as a dust fluid without stress and pressure in the non-relativistic limit after the Klein transformation; we consider the case using the post-Newtonian approximation. In the case of dark photon, the mass constraint may come from plausible cosmology and astrophysics.

Section \ref{sec:Proca} presents the massive photon electrodynamics and MHD in covariant forms. Section \ref{sec:Proca-ADM} presents the electrodynamics and MHD using the ADM formulation in generally curved spacetime. Section \ref{sec:Proca-cosmology} presents the electrodynamics and MHD in the context of perturbed Friedmann world model. Gauge transformation properties of EM fields and the vector potential are derived. Wave equations are presented assuming gravity is weakly varying compared with fields. Section \ref{sec:Proca-DM} derives the dark matter nature of the dark Proca field under the Klein transformation in non-relativistic limit. Section \ref{sec:Discussion} is a discussion.

We use $x^0 = \eta$ in the perturbation analysis in Sec.\ \ref{sec:Proca-cosmology}, and use $x^0 = ct$ in the post-Newtonian approximation in Sec.\ \ref{sec:Proca-DM}; we have $cdt = a d \eta$ with $a$ the cosmic scale factor. These allow us to use the related formulations in \cite{HN-2023-EM, Hwang-Noh-axion-2023}. When the photon has a mass, $c$ is no longer the speed of light. It can be considered as a constant (the relativistic limiting speed in vacuum) introduced by Einstein in his Special Theory of Relativity.

%
%
%
\section{Covariant formulation}
                                  \label{sec:Proca}

We consider EM fields with a potential $U$ of the vector potential $A_a$. As the EM part of the Lagrangian, in Heaviside-Lorentz units \cite{Jackson-1975}, we consider \cite{Ford-1989}
\bea
   & & {\cal L} = \sqrt{-g} \Big[ - {1 \over 4} F^{ab} F_{ab}
       + {1 \over c} J^a A_a - U (A^c A_c) \Big].
   \label{L}
\eea
For the Proca field, we have \cite{Proca}
\bea
   U = {1 \over 2} {m^2 c^2 \over \hbar^2} A^c A_c.
   \label{U}
\eea
Variation with respect to $g_{ab}$, using $J_a \propto 1/\sqrt{-g}$ \cite{Weinberg-1972}, gives
\bea
   T_{ab} = F_{ac} F_b^{\;\;c}
       - {1 \over 4} g_{ab} F^{cd} F_{cd}
       + 2 U^\prime A_a A_b - U g_{ab},
   \label{Tab}
\eea
where $U^\prime \equiv {\partial U \over \partial \xi}$ with $\xi \equiv A^c A_c$. Variation with respect to $A_a$ gives
\bea
   F^{ab}_{\;\;\;\; ;b}
       = {1 \over c} J^a - 2 U^\prime A^a.
   \label{M1}
\eea
The potential term in Eq.\ (\ref{M1}) can be absorbed as an effective four-current as
\bea
   {1 \over c} J_{\rm eff}^a = - 2 U^\prime A^a.
   \label{J-eff}
\eea
The other Maxwell's equation is the Bianchi identity
\bea
   \eta^{abcd} F_{bc,d} = 0,
   \label{M2}
\eea
with a solution $F_{ab} = \partial_a A_b - \partial_b A_a$. The vector potential $A_a$ is introduced in this way.

In the presence of the potential $U$, Eq.\ (\ref{M1}) together with the current conservation $J^a_{\;\; ;a} = 0$ {\it demands}
\bea
   (U^\prime A^a)_{;a} = 0,
   \label{Lorenz-cov}
\eea
which becomes the $A^a_{\;\;;a} = 0$ for a massive (Proca) field. For vanishing potential $U$, the Proca electrodynamics becomes Maxwell equations, and $A^a_{\;\;;a} = 0$ is the Lorenz gauge condition \cite{Jackson-1975, Jackson-Okun-2001}. In the presence of $U$ or the Proca mass term the vector potential acquires a physical significance and the gauge invariance is abandoned; we may call Eq.\ (\ref{Lorenz-cov}) the Lorenz condition.

Einstein's field equations in the covariant (1+3) formulation \cite{Ehlers-1993, Ellis-1971, Ellis-1973, Hawking-1966, Bertschinger-1995} are summarized in the Appendix A of \cite{HN-2023-EM}. The fluid quantities of the EM part needed for the equations are presented below, see Eq.\ (\ref{fluid-EM-cov}).

\subsection{Using EM fields}

The EM fields are defined using the observer's time-like four-vector $u_a$ as \cite{Ellis-1973}
\bea
   F_{ab} = u_a E_b - u_b E_a - \eta_{abcd} u^c B^d,
   \label{Fab-EB}
\eea
with $E_a u^a \equiv 0 \equiv B_a u^a$ and  $u^a u_a \equiv - 1$; here we consider $u_a$ to be a generic four-vector, and later we will specify it into the normal four-vector $n_a$ or the fluid four-vector $u_a$. We have $E_a = F_{ab} u^b$ and $B_a = F^*_{ab} u^b$; the dual tensor is defined as $F^*_{ab} = {1 \over 2} \eta_{abcd} F^{cd}$. The charge and current densities are defined using the same four-vector as
\bea
   J_a = \varrho c u_a + j_a, \quad j_a u^a \equiv 0.
   \label{J_a}
\eea
In Gaussian unit, we set $(F_{ab}, E_a, B_a, A_a) \rightarrow  (F_{ab}, E_a, B_a, A_a) / \sqrt{4 \pi}$ and $(J_a, j_a, \varrho) \rightarrow \sqrt{4 \pi} (J_a, j_a, \varrho)$.

Equations (\ref{M1}) and (\ref{M2}) give Maxwell's equations in terms of EM fields \cite{Ellis-1973}
\bea
   & & E^a_{\;\; ;b} h^b_a
       = \varrho_{\rm em}
       + 2 U^\prime A_a u^a
       - 2 \omega^a B_a,
   \label{Maxwell-cov-1} \\
   & & h^a_b \widetilde {\dot E}{}^b
       = \left( \eta^a_{\;\;bcd} u^d \omega^c
       + \sigma^a_{\;\;b}
       - {2 \over 3} \delta^a_b \theta \right) E^b
   \nonumber \\
   & & \qquad
       + \eta^{abcd} u_d
       \left( a_b B_c - B_{b;c} \right)
       - {1 \over c} j^{a}
       + 2 U^\prime h^b_a A_b,
   \label{Maxwell-cov-2} \\
   & & B^a_{\;\; ;b} h^b_a
       = 2 \omega^a E_a,
   \label{Maxwell-cov-3} \\
   & & h^a_b \widetilde {\dot B}{}^b
       = \left( \eta^a_{\;\;bcd} u^d \omega^c
       + \sigma^a_{\;\;b}
       - {2 \over 3} \delta^a_b \theta \right) B^b
   \nonumber \\
   & & \qquad
       - \eta^{abcd} u_d
       \left( a_b E_c - E_{b;c} \right),
   \label{Maxwell-cov-4}
\eea
where $\widetilde {\dot E}{}^a \equiv E^a_{\;\; ;b} u^b$; $\theta$, $\sigma_{ab}$ and $\omega_a$ are the expansion scalar, shear tensor and vorticity vector, respectively \cite{Ellis-1971, Ellis-1973, HN-2023-EM}.

Using Eqs.\ (\ref{J-eff}) and (\ref{J_a}), contributions from the vector potential can be absorbed using the effective charge and current densities as $\varrho \rightarrow \varrho + \varrho_{\rm eff}$ and $j_i \rightarrow j_i + j_i^{\rm eff}$ with
\bea
   \varrho_{\rm eff} = 2 U^\prime A_a u^a, \quad
       {1 \over c} j^{\rm eff}_a = - 2 U^\prime h^b_a A_b,
   \label{effective}
\eea
where $h_{ab} \equiv g_{ab} + u_a u_b$ is the spatial projection tensor orthogonal to $u_a$. In the following, our charge and current densities do not include the effective ones.

From Eq.\ (\ref{Tab}), we have
\bea
   & & \hskip -.9cm
       T_{ab}
       = {1 \over 2} ( E^2 + B^2 )
       ( u_a u_b + h_{ab} ) - E_a E_b - B_a B_b
   \nonumber \\
   & & \qquad
       \hskip -.9cm
       + ( u_a \eta_{bcde} + u_b \eta_{acde} ) E^c B^d u^e
       + 2 U^\prime A_a A_b - U g_{ab},
   \label{Tab-EM}
\eea
where $E^2 \equiv E^c E_c$, etc. The fluid quantities are measurable quantities and are associated with the observer's four-vector. The fluid quantities measured by an observer with a general four-vector $u_a$ are introduced as \cite{Ellis-1971}
\bea
   & & \hskip -1cm
       T_{ab} = \mu u_a u_b
       + p \left( g_{ab} + u_a u_b \right)
       + q_a u_b + q_b u_a + \pi_{ab},
   \label{Tab-fluid} \\
   & & \hskip -1cm
       \mu = T_{ab} u^a u^b, \quad
       p = {1 \over 3} T_{ab} h^{ab}, \quad
       q_a = - T_{cd} u^c h^d_a,
   \nonumber \\
   & & \hskip -1cm
       \pi_{ab} = T_{cd} h^c_a h^d_b - p h_{ab},
   \label{fluid-quantities}
\eea
where $\mu$, $p$, $q_a$ and $\pi_{ab}$ are the energy density, pressure, flux vector and anisotropic stress tensor, respectively. Using Eq.\ (\ref{Tab-EM}), we have
\bea
   & & \mu = {1 \over 2} ( E^2 + B^2 )
       + 2 U^\prime (A_a u^a)^2 + U,
   \nonumber \\
   & & p = {1 \over 6} ( E^2 + B^2 )
       + {2 \over 3} U^\prime [ (A_a u^a)^2 + A^2 ] - U,
   \nonumber \\
   & & q_a = \eta_{abcd} E^b B^c u^d
       - 2 U^\prime A_c u^c A_d h^d_a,
   \nonumber \\
   & & \pi_{ab}
       = - \Big[ E_a E_b + B_a B_b
       - {1 \over 3} h_{ab}
       \left( E^2 + B^2 \right) \Big]
   \nonumber \\
   & & \qquad
       + 2 U^\prime \Big(
       A_c h^c_a A_d h^d_b
       - {1 \over 3} A_c A_d h^{cd} h_{ab} \Big),
   \label{fluid-EM-cov}
\eea
where $A^2 \equiv A^c A_c$. The Poynting vector is identified as $S_i = c q_i$. The above fluid quantities are the ones based on $u_a$ for both fluid quantities and EM fields.

By decomposing $T_{ab}$ to the EM part (including A-part) as in Eq.\ (\ref{Tab-EM}) and the fluid (FL) part as in Eq.\ (\ref{Tab-fluid}), we have $T_{ab} = T_{ab}^{\rm FL} + T_{ab}^{\rm EM}$. The total energy-momentum tensor conservation demands $T^{ab}_{\;\;\;\; ;b} = 0$ and using Eq.\ (\ref{Tab}) for the EM part, we can show
\bea
   T^{ab}_{{\rm FL} ;b}
       = - T^{ab}_{{\rm EM} ;b}
       = {1 \over c} F^{ab} J_b.
   \label{Tab-consevation-FL}
\eea
Thus, the fluid conservation equations are not directly affected by the photon-mass; these are presented in Eqs.\ (43) and (44) of \cite{HN-2023-EM}.

Up to this point, the $u_a$ was a generic time-like four-vector. As mentioned, the EM fields are observable quantities and depend on the observer's four-vector. For $E_a$ and $B_a$ we will take the normal four-vector $u_a = n_a$, with $n_i \equiv 0$, associated with the Eulerian observer \cite{Smarr-York-1978, Wilson-Mathews-2003, Gourgoulhon-2012}. As we consider EM fields of the Eulerian observer, we have $u_a = n_a$ in Eqs.\ (\ref{Fab-EB})-(\ref{Tab-EM}). On the other hand, the fluid quantities in Eq.\ (\ref{fluid-quantities}) are often introduced by an observer comoving (at rest) with the fluid, and in that case we take $u_a$ to be the fluid four-vector; this applies to Eqs.\ (\ref{Tab-fluid})-(\ref{fluid-EM-cov}). For explicit forms of the two four-vectors, see below Eqs.\ (\ref{n_a-ADM}) and (\ref{u_a-ADM}). Relations of the EM fields and the charge and current densities measured by observers with the two four-vectors are presented in Eq.\ (45) of \cite{HN-2023-EM}.

\subsection{Using the vector potential}

The vector potential $A_a$ is defined as a solution of Eq.\ (\ref{M2}) and is not associated with the four-vector. In terms of the vector potential the EM fields, using the normal four-vector, are
\bea
   & & E_a \equiv F_{ab} n^b
       = A_{b;a} n^b - \widetilde {\dot A}_a,
   \nonumber \\
   & & B_a \equiv F^*_{ab} n^b
       = - \eta_{abcd} A^{b;c} n^d,
   \label{EB-A-cov}
\eea
where $\widetilde {\dot A}_a \equiv A_{a;b} n^b$. Using the vector potential, Eq.\ (\ref{M1}) gives
\bea
   \Box A_a - A^b_{\;\; ;ba} - R_{ab} A^b - 2 U^\prime A_a
       = - {1 \over c} J_a,
   \label{EOM-A}
\eea
and Eq.\ (\ref{M2}) is identically valid; for the Proca field, the second term vanishes due to the Lorenz condition.
Equation (\ref{Tab}) gives
\bea
   & & \hskip -.8cm
       T_{ab} = ( A_{a;c} - A_{c;a} )
       ( A_b^{\;\; ;c} - A^c_{\;\; ;b} )
   \nonumber \\
   & & \qquad \hskip -.8cm
       - {1 \over 2} g_{ab} A_{c;d} ( A^{c;d} - A^{d;c} )
       + 2 U^\prime A_a A_b - U g_{ab}.
   \label{Tab-A}
\eea
Above results are spacetime covariant.

\subsection{MHD approximation}
                                        \label{sec:MHD-cov}

Relativistic MHD with helical coupling was presented in Sec. VI of \cite{HN-2023-EM}. In the presence of the photon mass, we only need to replace the charge and current densities in the Gauss' and Amp$\grave{\rm e}$re's laws by adding the effective ones in Eq.\ (\ref{effective}); the charge and current densities in the following do not include the effective ones.

MHD considers a simple form of Ohm's law relating the current and electric field measured by a comoving (rest frame) observer of the fluid \cite{Weyl-1922, Eckart-1940, Jackson-1975}, i.e.,
\bea
   j^{(u)}_a = \sigma e_a,
   \label{Ohms-cov}
\eea
using $u_a$ as the fluid four-vector for both $j^{(u)}_a$ and $e_a$; $e_a \equiv E_a^{(u)}$ and $\sigma$ is the electric conductivity. As the MHD condition we further consider large conductivity compared with inverse of the characteristic time scale $T$ of the system, i.e., $\sigma \gg 1/T$ \cite{Somov-1994}. The {\it ideal} MHD considers $\sigma \rightarrow \infty$ limit with finite current, thus $e_a = 0$. Keeping finite $\sigma$ gives the {\it resistive} MHD.

The above Ohm's law is a simplified one often used in the MHD approximation to close the hydrodynamic equations coupled with Maxwell's equations, by specifying the electric field in Faraday's law. In general, the Ohm's law is derived by relating the charge current to velocity difference between ion and electron and using the momentum conservation equations of the ion and electron fluids. The simplified form in Eq.\ (24) follows by assuming negligible total charge density and $m_e \ll m_{ion}$, and demanding the oscillation frequency of the field negligible compared with the electron plasma frequency. Thus, the effective charge and current densities introduced in Eq.\ (15) do {\it not} affect the Ohm's law relating the current from the charge densities to the electric field in the comoving reference frame.

Using relations of electromagnetic quantities between comoving and normal four-vectors in Eq.\ (45) of \cite{HN-2023-EM}, Ohm's law in Eq.\ (\ref{Ohms-cov}) gives
\bea
   & & j_a + u_a j_b u^b
       + \varrho_{\rm em} c (n_a - \gamma u_a)
   \nonumber \\
   & & \qquad
       = \sigma \left( \gamma E_a + n_a E_b u^b
       - \eta_{abcd} u^b n^c B^d \right).
\eea
Gauss' law in Eq.\ (\ref{Maxwell-cov-1}), assuming the photon mass contribution is of the similar order as $\varrho_{\rm em}$, gives $E^a_{\;\;;b} h^b_a \sim E_a/L$ where $L$ is the characteristic length scale; we have $\omega_a = 0$ for the normal four-vector \cite{Ellis-1971}. Using the MHD condition, $\sigma \gg 1/T$, and $c (n_a - \gamma u_a) \sim L/T$, the convective current term, $\varrho_{\rm em} c (n_a - \gamma u_a)$ in the above equation can be ignored compared with $\sigma E_a$ term in the right-hand side. Thus, we have
\bea
   j_a + u_a j_b u^b
       = \sigma \left( \gamma E_a + n_a E_b u^b
       - \eta_{abcd} u^b n^c B^d \right).
   \label{Ohm-cov-MHD-1}
\eea
Now, we can set up the covariant MHD formulation.

Ohm's law in Eq.\ (\ref{Ohm-cov-MHD-1}) determines $E_a$
\bea
   \gamma E_a
       = \eta_{abcd} u^b n^c B^d
       + {1 \over \sigma} \left[ j_a
       + ( u_a - \gamma n_a) j_b u^b \right],
   \label{E-MHD-cov}
\eea
where we used $\gamma j_a u^a = \sigma E_a u^a$ which follows from Eq.\ (\ref{Ohm-cov-MHD-1}). We use Maxwell's equations in Eqs.\ (\ref{Maxwell-cov-1})-(\ref{Maxwell-cov-4}) using the normal four-vector $n_a$. Gauss' law in Eq.\ (\ref{Maxwell-cov-1}) determines $\varrho_{\rm em}$
\bea
   \varrho_{\rm em}
       = {1 \over 4 \pi} E^a_{\;\; ;b} h^b_a
       - 2 U^\prime A_a,
   \label{charge-MHD-cov}
\eea
where $h_{ab}$ is associated with $n_a$ and $\omega_{ab} = 0$ for the normal-frame. Amp$\grave{\rm e}$re's law in Eq.\ (\ref{Maxwell-cov-2}) determines $j^a$
\bea
   & & {1 \over c} j^{a}
       = \left( \sigma^a_{\;\;b}
       - {2 \over 3} \delta^a_b \theta \right) E^b
       + \eta^{abcd} n_d
       \left( a_b B_c - B_{b;c} \right)
   \nonumber \\
   & & \qquad
       - h^a_b \widetilde {\dot E}{}^b
       + 2 U^\prime h^b_a A_b,
   \label{j-MHD-cov}
\eea
where we kept the displacement current term ($\widetilde {\dot E}{}^b$-term), see below Eq.\ (161) of \cite{HN-2023-EM}.

Faraday's law in Eq.\ (\ref{Maxwell-cov-4}) gives the induction equation for $B^a$
\bea
   h^a_b \widetilde {\dot B}{}^b
       = \left( \sigma^a_{\;\;b}
       - {2 \over 3} \delta^a_b \theta \right) B^b
       - \eta^{abcd} n_d
       \left( a_b E_c - E_{b;c} \right),
   \label{Faraday-MHD-cov}
\eea
and the no-monopole condition in Eq.\ (\ref{Maxwell-cov-3}) gives
\bea
   B^a_{\;\; ;b} h^b_a
       = 0.
   \label{no-monopole-MHD-cov}
\eea
Equations (\ref{E-MHD-cov})-(\ref{no-monopole-MHD-cov}) provide Maxwell equations determining the magnetic field in MHD approximation. The EM variables $E_a$, $\varrho_{\rm em}$ and $j_a$ are determined by Eqs.\ (\ref{E-MHD-cov})-(\ref{j-MHD-cov}); $E_a$ and $j_a$ are coupled in Eqs.\ (\ref{E-MHD-cov}) and (\ref{j-MHD-cov}). The effect of photon-mass appears only in Eq.\ (\ref{j-MHD-cov}); Eq.\ (\ref{charge-MHD-cov}) is used to determine $\varrho_{\rm em}$. As shown below Eq.\ (\ref{Tab-consevation-FL}), the fluid conservation equations are not affected by the photon-mass, and are presented in Sec.\ II of \cite{HN-2023-EM}. In the following, if needed, we will indicate the covariant quantity by using a tilde.

%
%
%
\section{ADM formulation}
                                  \label{sec:Proca-ADM}

The ADM formulation is based on the following metric \cite{ADM, Bardeen-1980}
\bea
   g_{00} \equiv - N^2 + N^i N_i, \quad
       g_{0i} \equiv N_i, \quad
       g_{ij} \equiv \overline h_{ij},
   \label{ADM-metric-def}
\eea
where $\overline h{}^{ij}$ is an inverse of the three-space intrinsic metric $\overline h_{ij}$, and the index of $N_i$ is associated with $\overline h_{ij}$. We have
\bea
   \overline h_{ij} = g_{ij}, \quad
       N_i = g_{0i}, \quad
       N = 1/\sqrt{ - g^{00} }.
\eea

In the ADM notation, we set $\widetilde B_i \equiv \overline B_i$ and similarly for $\overline E_i$ and $\overline j_i$. For the vector potential, we set $\widetilde A_i \equiv \overline A_i$ and $\widetilde A_0 \equiv \overline A_0$. Indices of $\overline B_i$ etc. are associated with $\overline h_{ij}$. The normal four-vector is
\bea
   n_i \equiv 0, \quad
       n_0 = - N, \quad
       n^i = - {N^i \over N}, \quad
       n^0 = {1 \over N}.
   \label{n_a-ADM}
\eea
The fluid four-vector $u_a$ can be related to the normal four-vector as $u_a \equiv \gamma (n_a + v_a)$ where $v_a$, with $v_a n^a \equiv 0$, is the fluid velocity measured by an Eulerian observer associated with $n_a$. We can show
\bea
   & & u_i \equiv \gamma v_i, \quad
       u_0 = \gamma ( - N + N^i v_i ),
   \nonumber \\
   & & 
       u^i = \gamma \Big( - {N^i \over N}
       + \overline h^{ij} v_j \Big), \quad
       u^0 = {\gamma \over N},
   \label{u_a-ADM}
\eea
where the index of $v_i$ is associated with $g_{ab}$ as the metric. Using $v_i \equiv V_i$, with the index of $V_i$ associated with $\overline h_{ij}$ as the metric, we have
\bea
   v_i \equiv V_i, \quad
       v_0 = N_i V^i, \quad
       v^i = V^i, \quad
       v^0 = 0,
\eea
and we recover Eq.\ (55) in \cite{HN-2023-EM}; $\gamma = 1/\sqrt{1 - v^2}$ with $v^2 \equiv v_i v^i = V_i V^i$ is the Lorentz factor. The ADM formulation of Maxwell's equations needed for the derivation is presented in Sec.\ III and Appendix B of \cite{HN-2023-EM}.

Using the normal four-vector, Eq.\ (\ref{EB-A-cov}) gives
\bea
   & & \overline E_i = {1 \over N} ( \overline A_{0,i}
       - \overline A_{i,0} )
       + {N^j \over N} ( \overline A_{i,j}
       - \overline A_{j,i} ),
   \nonumber \\
   & & \overline B_i = \overline \eta_{ijk} \overline A{}^{k:j}.
   \label{EB-A-ADM}
\eea
From Eq.\ (\ref{effective}), we have
\bea
   \varrho_{\rm eff} = {2\over N} U^\prime ( \overline A_0
       - \overline A_i N^i ), \quad
       {1 \over c} \overline j{}^{\rm eff}_i = - 2 U^\prime \overline A{}^i.
   \label{effective-ADM}
\eea
For the Proca photon, we have $2 U^\prime = m^2 c^2/\hbar^2 = 1/\lambdabar^2$. Using these, Maxwell's equations in Eqs.\ (65)-(80) of \cite{HN-2023-EM} give
\bea
   & & \hskip -.8cm
       \overline E{}^i_{\; :i} = \varrho_{\rm em}
       + {2\over N} U^\prime ( \overline A_0
       - \overline A_i N^i ),
   \label{Maxwell-ADM-1} \\
   & & \hskip -.8cm
       \overline E{}^i_{\;,0}
       = N K \overline E{}^i
       + \overline E{}^i_{\; :j} N^j
       - \overline E{}^j N^i_{\; :j}
       + \overline \eta^{ijk} ( N \overline B_k )_{:j}
   \nonumber \\
   & & \qquad
       \hskip -.8cm
       - N \Big( {1 \over c} \overline j^i
       - 2 U^\prime \overline A{}^i \Big),
   \label{Maxwell-ADM-2} \\
   & & \hskip -.8cm
       \overline B{}^i_{\; :i} = 0,
   \label{Maxwell-ADM-3} \\
   & & \hskip -.8cm
       \overline B{}^i_{\;,0}
       = N K \overline B{}^i
       + \overline B{}^i_{\; :j} N^j
       - \overline B{}^j N^i_{\; :j}
       - \overline \eta^{ijk} ( N \overline E_k )_{:j},
   \label{Maxwell-ADM-4}
\eea
where a colon denotes the covariant derivative associated with the ADM metric $\overline h_{ij}$. The Lorenz condition in  Eq.\ (\ref{Lorenz-cov}) gives
\bea
   & & \hskip -.8cm
       {1 \over \sqrt{\overline h}} \Big[
       {\sqrt{\overline h} \over N} U^\prime
       ( - \overline A_0 + N^i \overline A_i ) \Big]_{,0}
   \nonumber \\
   & & \qquad
       \hskip -.8cm
       + \Big\{ {1 \over N} U^\prime \big[ N^i \overline A_0
       + ( N^2 \overline h^{ij} - N^i N^j ) \overline A_j \big]
       \Big\}_{|i} = 0.
   \label{Lorenz-ADM}
\eea
Using vector notation, we have
\bea
   & & \hskip -.5cm
       \overline \nabla \cdot \overline {\bf E}
       = \varrho_{\rm em}
       + {2\over N} U^\prime ( \overline A_0
       - \overline {\bf A} \cdot {\bf N} ),
   \label{M-ADM-1} \\
   & & \hskip -.5cm
       {1 \over \sqrt{\overline h}} ( \sqrt{\overline h}
       \overline {\bf E} )_{,0}
       = - \overline \nabla \times ( {\bf N} \times \overline {\bf E} )
       + {\bf N} \overline \nabla \cdot \overline {\bf E}
       + \overline \nabla \times ( N \overline {\bf B} )
   \nonumber \\
   & & \qquad
       \hskip -.5cm
       - N \Big( {1 \over c} \overline {\bf j}
       - 2 U^\prime \overline {\bf A} \Big),
   \label{M-ADM-2} \\
   & & \hskip -.5cm
       \overline \nabla \cdot \overline {\bf B} = 0,
   \label{M-ADM-3} \\
   & & \hskip -.5cm
       {1 \over \sqrt{\overline h}} ( \sqrt{\overline h}
       \overline {\bf B} )_{,0}
       = - \overline \nabla \times
       ( {\bf N} \times \overline {\bf B} )
       - \overline \nabla \times ( N \overline {\bf E} ),
   \label{M-ADM-4}
\eea
and
\bea
   \hskip -.2cm
   \overline {\bf E}
       = {1 \over N} \big[ \overline \nabla \overline A_0
       - \overline {\bf A}_{,0}
       - {\bf N} \times ( \overline \nabla \times
       \overline {\bf A} ) \big], \quad
       \overline {\bf B} = \overline \nabla \times
       \overline {\bf A},
   \label{EB-A-ADM-vector}
\eea
where $\overline {\bf B} \equiv \overline B{}^i$, etc., and the curl and divergence operators are associated with $\overline \eta_{ijk}$ and $\overline h_{ij}$; $\overline h \equiv {\rm det}(\overline h_{ij})$.

Using the vector potential, from Eqs.\ (\ref{Maxwell-ADM-1}) and (\ref{Maxwell-ADM-2}), using Eq.\ (\ref{EB-A-ADM}), we have
\bea
   & & \Big[ {1 \over N} ( \overline A_0{}^{|i}
       - \overline h{}^{ij} \overline A_{j,0} )
       + {N^j \over N} ( \overline A{}^i{}_{|j}
       - \overline A_j{}^{|i} ) \Big]_{|i}
   \nonumber \\
   & & \qquad
       = \varrho_{\rm em}
       + {2 \over N} U^\prime ( \overline A_0
       - \overline A_i N^i ),
   \label{A0-eq-ADM} \\
   & & {1 \over \sqrt{\overline h}} \Big\{ \sqrt{\overline h}
       \Big[ {1 \over N} ( \overline A_0{}^{|i}
       - \overline h{}^{ij} \overline A_{j,0} )
       + {N^j \over N} ( \overline A{}^i{}_{|j}
       - \overline A_j{}^{|i} ) \Big] \Big\}_{,0}
   \nonumber \\
   & & \qquad
       + \Big\{ {N^i \over N} ( \overline A_0{}^{|j}
       - \overline h{}^{jk} \overline A_{k,0} )
       - {N^j \over N} ( \overline A_0{}^{|i}
       - \overline h{}^{ik} \overline A_{k,0} )
   \nonumber \\
   & & \qquad
       + {N^k \over N} \big[ N^i ( \overline A{}^j{}_{|k}
       - \overline A_k{}^{|j} )
       - N^j ( \overline A{}^i{}_{|k}
       - \overline A_k{}^{|i} ) \big]
   \nonumber \\
   & & \qquad
       - N ( \overline A{}^{j|i} - \overline A{}^{i|j} ) \Big\}_{|j}
       = - N \Big( {1 \over c} \overline j{}^i
       - 2 U^\prime \overline A{}^i \Big)
   \nonumber \\
   & & \qquad
       + \Big[ \varrho_{\rm em}
       + { 2\over N} U^\prime ( \overline A_0
       - \overline A_j N^j ) \Big] N^i,
   \label{Ai-eq-ADM}
\eea
where we used properties of ADM quantities in \cite{HN-2023-EM}. Equations Eqs.\ (\ref{Maxwell-ADM-3}) and (\ref{Maxwell-ADM-4}) are identically valid.

Einstein's field equations in the ADM (3+1) formulation \cite{ADM, Bardeen-1980} are summarized in the Appendix B of \cite{HN-2023-EM}. The ADM fluid quantities are defined as
\bea
   & & E \equiv n_a n_b T^{ab},
       \quad
       J_i \equiv - n_b T^b_i, \quad
       S_{ij} \equiv T_{ij},
   \nonumber \\
   & &
       S \equiv \overline h{}^{ij} S_{ij}, \quad
       \overline S_{ij} \equiv S_{ij}
       - {1\over 3} \overline h_{ij} S.
   \label{ADM-fluid-def}
\eea
For the EM part, we have
\bea
   & & E
       = {1 \over 2} ( \overline E{}^2 + \overline B{}^2 )
       + 2 U^\prime ( \overline A_0 - N^i \overline A_i )^2 + U,
   \nonumber \\
   & & J_i
       = ( \overline {\bf E} \times \overline {\bf B} )_i
       - {1 \over N} 2 U^\prime ( \overline A_0
       - N^j \overline A_j ) \overline A_i,
   \nonumber \\
   & & S_{ij}
       = - \Big[ \overline E_i \overline E_j
       + \overline B_i \overline B_j
       - {1 \over 2} h_{ij} ( \overline E{}^2
       + \overline B{}^2 ) \Big]
   \nonumber \\
   & & \qquad
       + 2 U^\prime \overline A_i \overline A_j
       - U \overline h_{ij},
   \nonumber \\
   & & S
       = {1 \over 2} ( \overline E{}^2 + \overline B{}^2 )
       + 2 U^\prime \overline A{}^2 - 3 U,
   \label{ADM-fluid-EM}
\eea
where $\overline A^2 \equiv \overline A^i \overline A_i$, etc.

\subsection{Static fields}

We consider a static fields and, for simplicity, {\it ignore} $N_i$ as in weak gravity. Further {\it assuming} $\partial ({\rm metric}) \ll \partial ({\rm fields})$, Eq.\ (\ref{Lorenz-ADM}) gives $\overline A{}^i{}_{|i} = 0$, and Eqs.\ (\ref{A0-eq-ADM}) and (\ref{Ai-eq-ADM}) give
\bea
   ( \overline \Delta - 2 U^\prime ) \overline A_0
       = N \varrho_{\rm em}, \quad
       ( \overline \Delta - 2 U^\prime ) \overline {\bf A}
       = - {1 \over c} \overline {\bf j}.
   \label{A-eqs-N}
\eea

In the cosmological background, we have
\bea
   & & \hskip -.5cm
       N = a, \quad
       N_i = 0, \quad
       \overline h = a^6, \quad
       \overline h_{ij} = a^2 \delta_{ij}, \quad
       \overline A_0 \equiv a A_0,
   \nonumber \\
   & & \hskip -.5cm
       \overline B_i \equiv a B_i, \quad
       \overline E_i \equiv a E_i, \quad
       \overline j_i \equiv a j_i, \quad
       \overline A_i \equiv a A_i, \quad
   \nonumber \\
   & & \hskip -.5cm
       c \overline V_i \equiv a v_i, \quad
       \overline \eta_{ijk} = \sqrt{\overline h} \eta_{ijk}, \quad
       \overline \eta{}^{ijk} = {1 \over \sqrt{\overline h}} \eta_{ijk},
   \label{BG-quantities}
\eea
with indices of $B_i$, $E_i$, $j_i$, $A_i$, $v_i$, and $\eta_{ijk}$ associated with $\delta_{ij}$; in section \ref{sec:Proca-cosmology} we will consider the case in perturbed cosmology.

Eq.\ (\ref{A-eqs-N}) gives
\bea
   {\Delta \over a^2} A_0
       - 2 U^\prime A_0
       = \varrho_{\rm em}, \quad
       {\Delta \over a^2} {\bf A}
       - 2 U^\prime {\bf A}
       = - {1 \over c} {\bf j}.
   \label{A-eqs-a}
\eea
In a static background with $a \equiv 1$, for Proca photons, the solutions are \cite{Schrodinger-1943, Gintsburg-1964, Bicknell-1977}
\bea
   & & A_0 ({\bf x}) = - {1 \over 4 \pi } \int
       {e^{-|{\bf x} - {\bf x}^\prime|/\lambdabar} \over
       |{\bf x} - {\bf x}^\prime|}
       \varrho_{\rm em} ({\bf x}^\prime) d^3 x^\prime,
   \\
   & & {\bf A} ({\bf x}) = {1 \over 4 \pi c} \int
       {e^{-|{\bf x} - {\bf x}^\prime|/\lambdabar} \over
       |{\bf x} - {\bf x}^\prime|}
       {\bf j} ({\bf x}^\prime) d^3 x^\prime,
\eea
with ${\bf E} = \nabla A_0$ and ${\bf B} = \nabla \times {\bf A}$. Thus, EM fields and the vector potential decay exponentially beyond $\lambdabar$ away from the sources.

\subsection{MHD approximation}

Relativistic MHD with helical coupling in the ADM formulation is presented in Sec.\ IV.B of \cite{HN-2023-EM}. Here we present the case for massive photons. We are using the EM fields, charge and current densities defined in the normal four-vector.

Ohm's law in Eq.\ (\ref{Ohms-cov}), using Eqs.\ (60) and (61) of \cite{HN-2023-EM}, gives
\bea
   \overline {\bf j}
       + \gamma^2 {\bf V} {\bf V} \cdot \overline {\bf j}
       - \varrho_{\rm em} \gamma^2 c {\bf V}
       = \sigma \gamma \left( \overline {\bf E}
       + {\bf V} \times \overline {\bf B} \right),
   \label{Ohms-ADM}
\eea
where ${\bf V}$ is defined as $u_i \equiv \gamma V_i$ and is the fluid velocity measured by the Eulerian observer \cite{Banyuls-1997, HN-2023-EM}; the index of $V_i$ is associated with $\overline h_{ij}$. Using the MHD condition in Gauss' law in Eq.\ (\ref{M-ADM-1}), the convective current term, $c \varrho_{\rm em} \gamma^2 {\bf V}$, is negligible; we use $\nabla \cdot \overline {\bf E} \sim \overline {\bf E}/L$ and $c {\bf V} \sim L/T$, and assume the photon-mass contribution is negligible or of the similar order as $\varrho_{\rm em}$. Thus, Ohm's law in the MHD approximation gives
\bea
   \overline {\bf j}
       + \gamma^2 {\bf V} {\bf V} \cdot \overline {\bf j}
       = \sigma \gamma \left( \overline {\bf E}
       + {\bf V} \times \overline {\bf B} \right).
   \label{Ohms-ADM-SM}
\eea
For an ideal MHD, with $\sigma \rightarrow \infty$, we have
\bea
   \overline {\bf E}
       = - {\bf V} \times \overline {\bf B}.
\eea
Ohm's law in Eq.\ (\ref{Ohms-ADM-SM}) determines $\overline {\bf E}$ as
\bea
   \overline {\bf E}
       = - {\bf V} \times \overline {\bf B}
       + {1 \over \sigma \gamma} \left( \overline {\bf j}
       + \gamma^2 {\bf V} {\bf V} \cdot \overline {\bf j} \right).
   \label{E-MHD-ADM}
\eea
Gauss's law in Eq.\ (\ref{M-ADM-1}) determines $\varrho_{\rm em}$
\bea
   \varrho_{\rm em} = \overline E{}^i_{\; :i}
       - {2 \over N} U^\prime ( \overline A_0
       - \overline A_i N^i ).
   \label{charge-MHD-ADM}
\eea
Amp$\grave{\rm e}$re's law in Eq.\ (\ref{M-ADM-2}) determines $\overline {\bf j}$
\bea
   & & \hskip -.5cm
       \overline {\bf j} = {c \over N}
       \Big[
       - \overline \nabla \times ( {\bf N}
       \times \overline {\bf E} )
       + {\bf N} \overline \nabla \cdot \overline {\bf E}
       + \overline \nabla \times ( N \overline {\bf B} )
   \nonumber \\
   & & \qquad
       \hskip -.5cm
       - {1 \over \sqrt{\overline h}}
       ( \sqrt{\overline h} \overline {\bf E} )_{,0} \Big]
       + 2 c U^\prime \overline {\bf A},
   \label{j-MHD-ADM}
\eea
where the displacement current term (the last term inside parenthesis) can be ignored using the MHD condition. Faraday's law in Eq.\ (\ref{M-ADM-4}) gives
\bea
   {1 \over \sqrt{\overline h}}
       ( \sqrt{\overline h} \overline {\bf B} )_{,0}
       = - \overline \nabla \times
       ( {\bf N} \times \overline {\bf B} )
       - \overline \nabla \times ( N \overline {\bf E} ).
   \label{Faraday-MHD-ADM}
\eea
Equations (\ref{E-MHD-ADM})-(\ref{Faraday-MHD-ADM}) provide Maxwell equations determining the magnetic field in MHD approximation; $\overline {\bf E}$ and $\overline {\bf j}$ are still coupled in Eqs.\ (\ref{E-MHD-ADM}) and (\ref{j-MHD-ADM}).

We note that the above ADM formulation of electrodynamics and MHD are valid in general curved spacetime of Einstein's gravity. The other equations determining the gravity and fluid, are presented in \cite{HN-2023-EM}. The effect of photon-mass appears in Eqs.\ (\ref{charge-MHD-ADM}) and (\ref{j-MHD-ADM}); Eq.\ (\ref{charge-MHD-ADM}) is used to determine $\varrho_{\rm em}$. The fluid conservation equations are not affected by the photon-mass, and are presented in Sec.\ III of \cite{HN-2023-EM}.

We can use Eqs.\ (\ref{E-MHD-ADM}) and (\ref{j-MHD-ADM}) to remove $\overline {\bf E}$ and $\overline {\bf j}$ in Eq.\ (\ref{Faraday-MHD-ADM}). For simplicity, {\it ignoring} ${\bf N}$, like in the weak gravity, and {\it assuming} slow motion, thus setting $\gamma = 1$, we have the induction equation
\bea
   & & {1 \over \sqrt{\overline h}}
       ( \sqrt{\overline h} \overline {\bf B} )_{,0}
       - \overline \nabla \times
       ( N {\bf V} \times \overline {\bf B} )
   \nonumber \\
   & & \qquad
       + \overline \nabla \times \Big\{ {c \over \sigma}
       \big[ \overline \nabla \times
       ( N \overline {\bf B} )
       + 2 U^\prime N \overline {\bf A} \big] \Big\} = 0.
   \label{Induction-ADM}
\eea
For the Proca photon, ignoring perturbed order gravity (thus $N$ and $\overline h$ are spatially uniform) and assuming constant $\sigma$, we have
\bea
   & & \hskip -.4cm
       {1 \over N} {1 \over \sqrt{\overline h}}
       ( \sqrt{\overline h} \overline {\bf B} )_{,0}
       - \overline \nabla \times ( {\bf V} \times
       \overline {\bf B} )
       - {c \over \sigma} \Big( \overline \Delta
       \overline {\bf B}
       - {m^2 c^2 \over \hbar^2} \overline {\bf B} \Big)
   \nonumber \\
   & & \qquad
       \hskip -.4cm
       = 0.
   \label{Induction-uniform}
\eea

In the cosmological background, using Eq.\ (\ref{BG-quantities}), Eq.\ (\ref{Induction-uniform}) gives
\bea
   {1 \over a^2} ( a^2 {\bf B} )^{\displaystyle{\cdot}}
       - {1 \over a} \nabla \times ( {\bf v} \times {\bf B} )
       - {c^2 \over \sigma} \Big( {\Delta \over a^2} {\bf B}
       - {1 \over \lambdabar^2} {\bf B} \Big)
       = 0,
   \label{Induction-Friedmann}
\eea
and ${\bf B} = {1 \over a} \nabla \times {\bf A}$. In Minkowski background with $a \equiv 1$ this was derived in \cite{Byrne-1972, Ryutov-1997}. To the linear order, we can ignore the flux-freezing term and have a solution
\bea
   {\bf B} = {\bf B}_i {a_i^2 \over a^2} {\rm exp}
       \Big[ - {c^2 \over \sigma} \int_{t_i}^t
       \Big( {k^2 \over a^2} + {1 \over \lambdabar^2} \Big) dt \Big],
   \label{Solution-Friedmann}
\eea
where we set $\Delta = - k^2$. In the super-Compton scale, the photon-mass term causes exponential decay of the magnetic field ${\bf B} \propto e^{- (c^2/\sigma \lambdabar^2) t}/a^2$; $a^{-2}$ factor is due to dilution of the field due to expansion of the background. Consistency with the MHD predictions can be used to constrain the photon-mass using astrophysical observations of the scale of coherent magnetic field \cite{Ryutov-2007}.

%
%
%
\section{In perturbed cosmology}
                                  \label{sec:Proca-cosmology}

We consider a perturbed Friedmann world model with the metric
\bea
   & & g_{00} = - a^2 (1 + 2 \alpha), \quad
       g_{0i} = - a \chi_i,
   \nonumber \\
   & &
       g_{ij} = a^2 [ (1 + 2 \varphi ) \delta_{ij}
       + 2 C_{ij} ],
   \label{metric-pert}
\eea
where indices of $\chi_i$ an $C_{ij}$ are associated with the metric $\delta_{ij}$ and its inverse; $C_{ij}$ is transverse and tracefree with $C^j_{i,j} \equiv 0 \equiv C^i_i$; $\chi_i$ can be decomposed into the longitudinal (scalar-type) and transverse (vector-type) perturbations as $\chi_i = \chi_{,i} + \chi^{(v)}_i$ with $\chi^{(v)i}_{\;\;\;\;\;,i} \equiv 0$; $x^0 = \eta$. We took spatial gauge conditions which leave all perturbation variables spatially gauge-invariant to the linear order, and the temporal gauge condition is not imposed yet \cite{Bardeen-1988}; for the gauge transformation properties, see below. The inverse metric is
\bea
   & & g^{00} = - {1 \over a^2} (1 - 2 \alpha), \quad
       g^{0i} = - {1 \over a^3} \chi^i,
   \nonumber \\
   & &
       g^{ij} = {1 \over a^2} [ (1 - 2 \varphi ) \delta^{ij}
       - 2 C^{ij} ].
\eea

In the following we consider a massive Proca field, thus $2 U^\prime = m^2 c^2/\hbar^2 = 1/\lambdabar^2$. Maxwell's equations in the ADM formulation are given in Eqs.\ (\ref{Maxwell-ADM-1})-(\ref{Maxwell-ADM-4}). The ADM quantities are
\bea
   & & N = a (1 + \alpha), \quad
       N_i = - a \chi_i, \quad
       N^i = - {1 \over a} \chi^i,
   \nonumber \\
   & &
       \overline h_{ij} = a^2 [ (1 + 2 \varphi) \delta_{ij}
       + 2 C_{ij} ], \quad
       {\rm det}(\overline h_{ij})
       = a^6 (1 + 6 \varphi),
   \nonumber \\
   & & \overline h{}^{ij} = {1 \over a^2}
       [ (1 - 2 \varphi) \delta^{ij} - 2 C^{ij} ],
   \nonumber \\
   & &
       K = - {1 \over a} \Big[
       3 {a^\prime \over a} ( 1 - \alpha )
       + 3 \varphi^\prime + {\Delta \over a} \chi \Big].
\eea
The normal four-vector is
\bea
   & & n_i \equiv 0, \quad
       n_0 = - a (1 + \alpha),
   \nonumber \\
   & &
       n^i = {1 \over a^2} \chi^i, \quad
       n^0 = {1 \over a} (1 - \alpha).
\eea

We introduce
\bea
   \widetilde B_i \equiv a B_i, \quad
       \widetilde A_i \equiv a A_i, \quad
       \widetilde A_0 \equiv a A_0,
   \label{EB-cov-pert}
\eea
and similarly for $E_i$ and $j_i$; indices of $B_i$, etc. are associated with $\delta_{ij}$ and its inverse. Using $n^a \widetilde B_a \equiv 0$, we have
\bea
   & & \widetilde B_0 = - B_i \chi^i, \quad
       \widetilde B{}^i = {1 \over a} [ ( 1 - 2 \varphi ) B^i
       - 2 C^{ij} B_j ],
   \nonumber \\
   & & \widetilde B{}^0 = 0,
\eea
and similarly for $\widetilde E_a$ and $\widetilde j_a$. In the ADM notation, we have
\bea
   \overline B_i \equiv a B_i, \quad
       \overline B^i \equiv \overline h{}^{ij} \overline B_j
       = {1 \over a} [ (1 - 2 \varphi) \delta^{ij}
       - 2 C^{ij} ] B_j,
\eea
and similarly for $\overline E_i$, $\overline j_i$, and $\overline A_i$.

From Eqs.\ (\ref{effective}), we have
\bea
   & & \varrho_{\rm eff}
       = {m^2 c^2 \over \hbar^2} \Big[ ( 1 - \alpha ) A_0
       + {1 \over a} A_i \chi^i \Big],
   \nonumber \\
   & &
       {1 \over c} j_i^{\rm eff}
       = - {m^2 c^2 \over \hbar^2} A_i.
   \label{effective-Proca}
\eea

\begin{widetext}
\subsection{Maxwell equations}

From Eq.\ (\ref{EB-A-cov}), we have
\bea
   E_i = {1 \over a} \Big[ (1 - \alpha) \Big( A_{0,i}
       - A_i^\prime - {a^\prime \over a} A_i \Big)
       + {1 \over a} ( A_{j,i} - A_{i,j} ) \chi^j \Big], \quad
       B_i = {1 \over a} [ ( 1 - \varphi ) \delta_{ij}
       + 2 C_{ij} ]  \eta^{jk\ell} \partial_k A_\ell.
   \label{EB-A}
\eea

Maxwell's equations give
\bea
   & & {1 \over a} \{ [ ( 1 + \varphi ) \delta^{ij}
       - 2 C^{ij} ] E_j \}_{,i}
       = (1 + 3 \varphi) \varrho
       + {m^2 c^2 \over \hbar^2}
       \Big[ (1 - \alpha + 3 \varphi) A_0
       + {1 \over a} A_i \chi^i \Big],
   \label{M-1} \\
   & & {1 \over a^3} \{ a^2 [ ( 1 + \varphi ) \delta^{ij}
       - 2 C^{ij} ] E_j \}^\prime
       - {1 \over a} \eta^{ijk} \partial_j
       \Big[ ( 1 + \alpha ) B_k
       - {1 \over a} \eta_{k\ell m} E^\ell \chi^m \Big]
   \nonumber \\
   & & \qquad
       = - [ ( 1 + \alpha + \varphi ) \delta^{ij}
       - 2 C^{ij} ] \Big( {1 \over c} j_j
       - {m^2 c^2 \over \hbar^2} A_j \Big)
       - {1 \over a} \Big( \varrho
       + {m^2 c^2 \over \hbar^2} A_0 \Big) \chi^i,
   \label{M-2} \\
   & & {1 \over a} \{ [ ( 1 + \varphi ) \delta^{ij}
       - 2 C^{ij} ] B_j \}_{,i}
       = 0,
   \label{M-3} \\
   & & {1 \over a^3} \{ a^2 [ ( 1 + \varphi ) \delta^{ij}
       - 2 C^{ij} ] B_j \}^\prime
       + {1 \over a} \eta^{ijk} \partial_j
       \Big[ ( 1 + \alpha ) E_k
       + {1 \over a} \eta_{k\ell m} B^\ell \chi^m \Big]
       = 0,
   \label{M-4}
\eea
where a prime indicates the time derivative associated with the conformal time $\eta$. In terms of $A_a$, using Eq.\ (\ref{EB-A}), Eqs.\ (\ref{M-1}) and (\ref{M-2}) give
\bea
   & & {1 \over a^2} \Big\{
       [ ( 1 - \alpha + \varphi ) \delta^{ij}
       - 2 C^{ij} ] \Big( A_{0,j}
       - A_j^\prime - {a^\prime \over a} A_j \Big)
       + {1 \over a} \Big( A^{j,i} - A^{i,j} \Big) \chi_j \Big\}_{,i}
   \nonumber \\
   & & \qquad
       = (1 + 3 \varphi) \varrho
       + {m^2 c^2 \over \hbar^2}
       \Big[ ( 1 - \alpha + 3 \varphi) A_0
       + {1 \over a} A_i \chi^i \Big],
   \label{A-0} \\
   & & {1 \over a^3} \Big\{ a
       \Big[ [ ( 1 - \alpha + \varphi ) \delta^{ij} - 2 C^{ij} ]
       \Big( A_j^\prime + {a^\prime \over a} A_j
       - A_{0,j} \Big)
       + {1 \over a} ( A^{i,j} - A^{j,i} ) \chi_j \Big] \Big\}^\prime
       + {1 \over a^2} \partial_j \Big[
       ( 1 + \alpha - \varphi ) ( A^{j,i} - A^{i,j} )
   \nonumber \\
   & & \qquad
       + 2 C^i_k ( A^{k,j} - A^{j,k} )
       - 2 C^j_k ( A^{k,i} - A^{i,k} )
       + \Big( A^{i\prime} + {a^\prime \over a} A^i
       - A_0^{\;,i} \Big) {1 \over a} \chi^j
       - \Big( A^{j\prime} + {a^\prime \over a} A^j
       - A_0^{\;,j} \Big) {1 \over a} \chi^i \Big]
   \nonumber \\
   & & \qquad
       = [ ( 1 + \alpha + \varphi ) \delta^{ij}
       - 2 C^{ij} ] \Big( {1 \over c} j_j
       - {m^2 c^2 \over \hbar^2} A_j \Big)
       + \Big( \varrho
       + {m^2 c^2 \over \hbar^2} A_0 \Big) {1 \over a} \chi^i,
   \label{A-i}
\eea
respectively, and  Eqs.\ (\ref{M-3}) and (\ref{M-4}) are identically valid. The Lorenz condition in Eq.\ (\ref{Lorenz-cov}) gives a constraint
\bea
   {1 \over a^3} \Big\{ a^3 \Big[
       ( 1 - \alpha + 3 \varphi ) A_0
       + {1 \over a} \chi^i A_i \Big] \Big\}^\prime
       = \Big\{ [ ( 1 + \alpha + \varphi ) \delta^{ij}
       - 2 C^{ij} ] A_j
       - {1 \over a} \chi^i A_0 \Big\}_{,i}.
   \label{Lorenz}
\eea
We can check that, Eq.\ (\ref{Lorenz}) is consistent with Eqs.\ (\ref{A-0}) and (\ref{A-i}); we use the four-current conservation $J^a_{\;\; ;a} = 0$ which gives
\bea
   {1 \over a^3} [ a^3 (1 + 3 \varphi) \varrho ]^\prime
       + \Big\{ [ (1 + \alpha + \varphi) \delta^{ij}
       - 2 C^{ij} ] {1 \over c} j_j
       + {1 \over a} \varrho \chi^i \Big\}_{,i}
       = 0.
   \label{current-conserv}
\eea
The above equations are valid without imposing the temporal gauge condition.

Assuming homogeneous fields, thus ignoring the metric perturbations, and ignoring the charge and current densities, Eqs.\ (\ref{M-1}) and (\ref{A-0}) give $\bar A_0 = 0$; a bar indicates the homogeneous background quantity. Equations (\ref{EB-A}), (\ref{M-2}) and (\ref{A-i}) give
\bea
   \bar E_i = - {1 \over a^2}
       ( a \bar A_i)^\prime, \quad
       \bar B_i = 0, \quad
       {1 \over a^3} ( a^2 \bar E_i )^\prime
       = - {1 \over a^3} (a \bar A_i )^{\prime\prime}
       = {m^2 c^2 \over \hbar^2} \bar A_i.
   \label{BG-eqs}
\eea

\subsection{MHD approximation}

The induction equation of MHD in the cosmological background is derived in Eq.\ (\ref{Induction-Friedmann}). Now, we consider a linearly perturbed cosmological background, and assume the spatial and temporal derivatives of the metric perturbations are negligible compared with that of the EM field, i.e., $\partial_a {\rm (metric \; perturbation)} \ll \partial_a {\rm (fields)}$. The Ohm's law in Eq.\ (\ref{Ohms-ADM-SM}) gives
\bea
   {\bf j} = \sigma \Big( {\bf E}
       + {1 \over c} {\bf v} \times {\bf B} \Big).
\eea
The Amp$\grave{\rm e}$re's law in Eq.\ (\ref{j-MHD-ADM}) gives
\bea
   {1 \over c} j^i
       = {1 \over a} \big[ ( 1 - \varphi ) \eta^{ijk}
       + 2 C_\ell^i \eta^{\ell jk} \big] \nabla_j B_k
       - {1 \over a^2} \chi_j \nabla^j E^i
       + {m^2 c^2 \over \hbar^2} A^i.
\eea
The Faraday's law in Eq.\ (\ref{Faraday-MHD-ADM}) gives
\bea
   {1 \over a^2} ( a^2 B^i )^{\displaystyle{\cdot}}
       = - {c \over a} \big[ ( 1 + \alpha - \varphi ) \eta^{ijk}
       + 2 C_\ell^i \eta^{\ell jk} \big] \nabla_j E_k
       - {c \over a^2} \chi_j \nabla^j B^i.
\eea
Replacing ${\bf E}$ and ${\bf j}$ using the Ohm's law and the Amp$\grave{\rm e}$re's law, respectively, using Eq.\ (\ref{EB-A}) and assuming constant $\sigma$, we have the induction equation
\bea
   {1 \over a^2} ( a^2 B^i )^{\displaystyle{\cdot}}
       = {1 \over a} \big[ \nabla \times
       ( {\bf v} \times {\bf B} ) \big]^i
       + {c^2 \over \sigma} \Big[
       ( 1 + \alpha - 2 \varphi ) {\Delta \over a^2} B^i
       - {2 \over a^2} \big( C^{jk} \nabla_j \nabla_k B^i
       - C^k_j \nabla_k \nabla^i B^j \big)
       - {m^2 c^2 \over \hbar^2} ( 1 + \alpha ) B^i \Big],
   \label{Induction-pert}
\eea
where we ignored $\chi_i$ (by imposing the zero-shear gauge and ignoring the vector-type perturbation, see later). This also follows directly from Eq.\ (\ref{Induction-ADM}) valid in general curved spacetime. Ignoring perturbations, Eq.\ (\ref{Induction-pert}) becomes Eq.\ (\ref{Induction-Friedmann}). The fluid conservation equations are not affected by the photon-mass, and are presented in Sec.\ IV of \cite{HN-2023-EM}.

\subsection{Gauge transformation}

Under a gauge (infinitesimal coordinate) transformation, $\widehat x^a = x^a + \xi^a$, using
\bea
   \phi (x^e) = \widehat \phi (\widehat x^e), \quad
       g_{ab} (x^e) = {\partial \widehat x^c \over \partial x^a}
       {\partial \widehat x^d \over \partial x^b}
       \widehat g_{cd} (\widehat x^e), \quad
       F_{ab} (x^e) = {\partial \widehat x^c \over \partial x^a}
       {\partial \widehat x^d \over \partial x^b}
       \widehat F_{cd} (\widehat x^e), \quad
       J_a (x^e) = {\partial \widehat x^b \over \partial x^a}
       \widehat J_b (\widehat x^e),
\eea
at the same spacetime point $x^e$, to the linear order, we have
\bea
   & & \widehat \phi = \phi - \phi_{,a} \xi^a, \quad
       \widehat g_{ab} = g_{ab} - g_{ab,c} \xi^c
       - g_{ac} \xi^c_{\;,b} - g_{bc} \xi^c_{\;,a},
   \nonumber \\
   & &
       \widehat F_{ab} = F_{ab} - F_{ab,c} \xi^c
       - F_{ac} \xi^c_{\;,b} - F_{cb} \xi^c_{\;,a}, \quad
       \widehat J_a = J_a - J_{a,b} \xi^b - J_b \xi^b_{\;,a}.
   \label{GT-Fab}
\eea
We consider the most general perturbations in the flat Friedmann world model with the metric
\bea
   g_{00} = - a^2 ( 1 + 2 \alpha ), \quad
       g_{0i} = - a^2 ( \beta_{,i} + B^{(v)}_i ), \quad
       g_{ij} = a^2 [ ( 1 + 2 \varphi ) \delta_{ij}
       + 2 \gamma_{,ij} + C_{i,j} + C_{j,i} + 2 C_{ij} ],
   \label{metric-general}
\eea
where $B^{(v)}_i$ and $C_i$ are transverse with $B^{(v)i}_{\;\;\;\;\;\;,i} = 0 = C^i_{\;,i}$; indices of $B^{(v)}_i$ and $C_i$ are associated with $\delta_{ij}$ and its inverse.

Under the gauge transformation, we have
\bea
   & & \widehat \alpha = \alpha - {1 \over a} (a \xi^0)^\prime,
       \quad
       \widehat \beta = \beta - \xi^0
       + \Big( {1 \over a} \xi \Big)^\prime, \quad
       \widehat \gamma = \gamma - {1 \over a} \xi, \quad
       \widehat \varphi = \varphi - {a^\prime \over a} \xi^0, \quad
       \widehat \chi_i = \chi_i - a \xi^0_{\;,i},
   \nonumber \\
   & & \widehat B^{(v)}_i = B^{(v)}_i + \xi^{(v)\prime}_i, \quad
       \widehat C_i = C_i - \xi^{(v)}_i, \quad
       \widehat C_{ij} = C_{ij},
\eea
where we set $\xi_i \equiv {1 \over a} \xi_{,i} + \xi^{(v)}_i$. We defined $\chi_i \equiv a [ ( \beta + \gamma^\prime )_{,i} + B^{(v)}_i + C^{(v)\prime}_i ] \equiv \chi_{,i} + \chi^{(v)}_i$ which is a unique spatially gauge invariant combination; by using $\chi$ and $\chi^{(v)}_i$ instead of $\beta$, $\gamma$, $B^{(v)}_i$, and $C_i$, all remaining variables are spatially gauge invariant \cite{Bardeen-1988}. Equivalently, in the spatial gauge conditions setting $\gamma = 0 = C_i$, we have $\chi_i = a ( \beta_{,i} + B^{(v)}_i) $, and all the perturbation variable are spatially gauge invariant.

In Eq.\ (\ref{metric-pert}) we already imposed these spatial gauge conditions without losing any generality \cite{Bardeen-1988}. But the temporal gauge condition is not imposed which can be used for our convenience depending on the situation. As the temporal gauge condition we can set $\varphi = 0$ (uniform-curvature gauge), $\chi = 0$ (zero-shear gauge) or others in all coordinates; each of these conditions removes the gauge degree of freedom completely, and the remaining variables are (spatially and temporally) gauge invariant. This statement on gauge invariance can be continued to fully nonlinear order in cosmological perturbations \cite{Noh-Hwang-2004, Hwang-Noh-2013}. Whereas, $\alpha = 0$ (synchronous gauge) is also a legitimate gauge condition, but imposing this in all coordinates still leaves $\xi^0 ({\bf x}, x^0) \propto 1/a$, which is the remnant gauge mode.

The EM fields are observable quantities, and are introduced by projection of $F_{ab}$ using the observer's four-vector \cite{Ellis-1973}. Using the normal four-vector, we have
\bea
   & & F_{0i} = - a^2 [ ( 1 + \alpha ) E_i
       + \eta_{ijk} B^j ( \beta^{,k} + B^{(v)k} ) ],
   \nonumber \\
   & & F_{ij} = a^2 \eta_{ijk} [ ( 1 + \varphi
       + \Delta \gamma ) B^k
       - 2 \gamma^{,k\ell} B_\ell
       - ( C^{k,\ell} + C^{\ell,k} ) B_\ell
       - 2 C^{k\ell} B_\ell ],
   \nonumber \\
   & & J_0 = - a [ ( 1 + \alpha ) \varrho c
       + ( \beta^{,i} + B^{(v)i} ) j_i ], \quad
       J_i = a j_i.
   \label{GT-Fab-EB}
\eea
Using Eqs.\ (\ref{GT-Fab}) and (\ref{GT-Fab-EB}), we have
\bea
   & & \widehat E_i = E_i - {1 \over a} ( a E_i )^\prime \xi^0
       + \eta_{ijk} B^j \xi^{0,k}
       - E_{i,j} \xi^j - E_j \xi^j_{\;,i}, \quad
       \widehat B_i = B_i - {1 \over a} ( a B_i )^\prime \xi^0
       - \eta_{ijk} E^j \xi^{0,k}
       - B_{i,j} \xi^j - B_j \xi^j_{\;,i},
   \nonumber \\
   & & \widehat \varrho = \varrho - \varrho^\prime \xi^0
       - \varrho_{,i} \xi^i
       + {1 \over c} j_i \xi^{0,i}, \quad
       \widehat j_i = j_i - {1 \over a} ( a j_i )^\prime \xi^0
       + \varrho c \xi^0_{\;,i}
       - j_{i,j} \xi^j - j_j \xi^j_{\;,i}.
   \label{GT-EB}
\eea
Notice that the temporal gauge transformation causes mixing between electric and magnetic fields, and between charge and current densities.

As derivatives of vector potential are related to the field strength tensor and EM fields, it is nontrivial to read the gauge transformation of the vector potential out of the transformations of latter ones. Instead, we can use the inhomogeneous Maxwell's equations for the purpose. Although the vector potential appears in the equation only for the massive case, the derived results apply generally. In the main text, we imposed $\gamma \equiv 0 \equiv C_i$ as the spatial gauge condition, and we have $\xi = 0 = \xi^{(v)}_i$, thus $\xi_i = 0$; the spatial gauge degree of freedom is completely fixed and the remaining perturbation variables are equivalently spatially gauge invariant. From Eqs.\ (\ref{M-1}) and (\ref{M-2}), using Eq.\ (\ref{GT-EB}), we have
\bea
   \widehat A_0 = A_0
       - a^2 A_0 \Big( {1 \over a^2} \xi^0 \Big)^\prime
       - A^i_{\;,i} \xi^0  - A_{0,i} \xi^i
       - A_i \xi^{i\prime}, \quad
       \widehat A_i = A_i - {1 \over a} ( a A_i )^\prime \xi^0
       - A_0 \xi^0_{\;,i} - A^j \xi_{j,i} - A_{i,j} \xi^j,
\eea
where we recovered the spatial gauge transformation by deriving the two inhomogeneous Maxwell's equations under the general metric in Eq.\ (\ref{metric-general}). By setting $\xi_i \equiv 0$, we can show that Eqs.\ (\ref{M-1})-(\ref{current-conserv}) are gauge invariant.

\subsection{Wave equations}

Assuming spatial and temporal derivatives of the metric (including the background) are negligible compared with that of the EM field, i.e., $\partial_a {\rm (metric)} \ll \partial_a {\rm (fields)}$, we can derive wave equations for EM fields in closed forms. From Eqs.\ (\ref{M-1})-(\ref{M-4}), using Eq.\ (\ref{EB-A}), we have
\bea
   & & {1 \over a^2} E_i^{\prime\prime}
       + {2 \over a^3} E^\prime_{i,j} \chi^j
       + \Big[ ( 1 + 2 \alpha ) {m^2 c^2 \over \hbar^2}
       - ( 1 + 2 \alpha - 2 \varphi ) {\Delta \over a^2}
       \Big] E_i
       + {2 \over a^2} E_{i,jk} C^{jk}
   \nonumber \\
   & & \qquad
       = - {1 \over a} \Big[ (1 + 2 \alpha) \varrho_{,i}
       + ( 1 + \alpha ) {1 \over c} j_i^\prime
       + {1 \over c a} j_{i,j} \chi^j \Big],
   \\
   & & {1 \over a^2} B_i^{\prime\prime}
       + {2 \over a^3} B^\prime_{i,j} \chi^j
       + \Big[ ( 1 + 2 \alpha ) {m^2 c^2 \over \hbar^2}
       - ( 1 + 2 \alpha - 2 \varphi ) {\Delta \over a^2}
       \Big] B_i
       + {2 \over a^2} B_{i,jk} C^{jk}
   \nonumber \\
   & & \qquad
       = {1 \over a c} \big[ (1 + 2 \alpha - \varphi)
       \eta_{ijk} \partial^j j^k
       + 2 C_i^j \eta_{jk\ell} \partial^k j^\ell \big].
\eea

In terms of the vector potential, Eqs.\ (\ref{A-0})-(\ref{Lorenz}) give
\bea
   & & {1 \over a^2} A_0^{\prime\prime}
       + {2 \over a^3} A_{0,i}^\prime \chi^i
       + \Big[ ( 1 + 2 \alpha ) {m^2 c^2 \over \hbar^2}
       - ( 1 + 2 \alpha - 2 \varphi ) {\Delta \over a^2}
       \Big] A_0
       + {2 \over a^2} A_{0,ij} C^{ij}
       = - ( 1 + 3 \alpha ) \varrho
       - {1 \over a c} j_i \chi^i,
   \\
   & & {1 \over a^2} A^{i\prime\prime}_{\;,i}
       + {2 \over a^3} A^{i\prime}_{\;,ij} \chi^j
       + \Big[ ( 1 + 2 \alpha ) {m^2 c^2 \over \hbar^2}
       - ( 1 + 2 \alpha - 2 \varphi ) {\Delta \over a^2}
       \Big] A^i_{\;,i}
       + {2 \over a^2} A^i_{\;,ijk} C^{jk}
       = (1 + 2 \alpha) {1 \over c} j^i_{\;,i},
   \\
   & & {1 \over a^2} \eta^{ijk} \partial_j A_k^{\prime\prime}
       + {2 \over a^3} \eta^{ijk} \partial_j A_{k,\ell}^\prime \chi^\ell
       + \Big[ ( 1 + 2 \alpha ) {m^2 c^2 \over \hbar^2}
       - ( 1 + 2 \alpha - 2 \varphi ) {\Delta \over a^2}
       \Big]  \eta^{ijk} \partial_j A_k
       + {2 \over a^2}  \eta^{ijk}
       \partial_j A_{k,\ell m} C^{\ell m}
   \nonumber \\
   & & \qquad
       = ( 1 + 2 \alpha ) {1 \over c}  \eta^{ijk} \partial_j j_k.
\eea
Thus, in this case all the wave equations are in closed forms.

Ignoring charge and current densities, we have
\bea
   {1 \over a^2} B_i^{\prime\prime}
       + {2 \over a^3} B^\prime_{i,j} \chi^j
       + \Big[ ( 1 + 2 \alpha ) {m^2 c^2 \over \hbar^2}
       - ( 1 + 2 \alpha - 2 \varphi ) {\Delta \over a^2}
       \Big] B_i
       + {2 \over a^2} B_{i,jk} C^{jk} = 0,
\eea
and {\it the same} equation for $B_i$ applies to $E_i$, $A_0$, $A^i_{\;,i}$, and $\eta^{ijk} \partial_j A_k$. The scalar ($\alpha$, $\varphi$ and $\chi$), vector ($\chi_i^{(v)}$) and tensor ($C_{ij}$) type perturbations affect the wave equations. For the scalar perturbation we still have a freedom to impose the temporal gauge condition. Zero-shear gauge takes $\chi \equiv 0$, and $\alpha$ and $- \varphi$ can be interpreted as the Newtonian and post-Newtonian gravitational potential. For vanishing anisotropic stress, we have $\varphi = - \alpha$. In Minkowski background, each component of these equations is Klein-Gordon equation \cite{Gintsburg-1964}.

The gravity affects the index of refraction as $n = 1 - \alpha + \varphi$ and the photon mass term is increased by ($1 + 2 \alpha$) factor. The vector perturbation appears in the damping part of the wave equation. Without a source for rotational perturbation, in expanding stage we have $\chi_i^{(v)} \propto 1/a$ \cite{Bardeen-1980}; the frame-dragging potential ${\tt \Psi}$ in \cite{Bardeen-1980} is proportional to our $\chi_i^{(v)}/a$. The tensor perturbation (gravitational waves) appears in a second order spatial derivative term. 
In this limit of ignoring derivatives of the metric compared with those of the field, we do not have the rotation of polarization plane due to gravity.

\subsection{Ignoring metric perturbations}

In the Friedmann background, thus ignoring metric perturbations, Eqs.\ (\ref{M-1})-(\ref{M-4}) give
\bea
   & & {1 \over a} E^i_{\;,i}
       = {m^2 c^2 \over \hbar^2} A_0 + \varrho,
   \label{M-Friedmann-1} \\
   & & {1 \over a^3} ( a^2 E^i )^\prime
       - {1 \over a} \eta^{ijk} \partial_j B_k
       = {m^2 c^2 \over \hbar^2} A^i - {1 \over c} j^i,
   \label{M-Friedmann-2} \\
   & & {1 \over a} B^i_{\;\;,i} = 0,
   \label{M-Friedmann-3} \\
   & & {1 \over a^3} ( a^2 B^i )^\prime
       + {1 \over a} \eta^{ijk} \partial_j E_k
       = 0.
   \label{M-Friedmann-4}
\eea
In terms of the vector potential, Eqs.\ (\ref{A-0})-(\ref{current-conserv}) give
\bea
   & & {1 \over a^2} \Big( A^{i\prime}
       + {a^\prime \over a} A^i \Big)_{,i}
       = \Big( {\Delta \over a^2}
       - {m^2 c^2 \over \hbar^2} \Big) A_0
       - \varrho,
   \label{A-0-Friedmann} \\
   & & {1 \over a^3} \Big[ a \Big(
       A_i^\prime + {a^\prime \over a} A_i - A_{0,i} \Big)
       \Big]^\prime
       + {1 \over a^2} ( A^j_{\;,ji}
       - \Delta A_i )
       = - {m^2 c^2 \over \hbar^2} A_i
       + {1 \over c} j_i,
   \label{A-i-Friedmann} \\
   & & A^i_{\;,i}
       = {1 \over a^3} ( a^3 A_0 )^\prime,
   \label{Lorenz-Friedmann} \\
   & & {1 \over a^3} ( a^3 \varrho )^\prime
       + {1 \over c} j^i_{\;,i}
       = 0,
   \label{current-conserv-Friedmann}
\eea
where the third equation is the Lorenz condition valid for non-vanishing photon mass. Equation (\ref{EB-A}) gives the relation
\bea
   E_i = {1 \over a} \Big( A_{0,i}
       - A_i^\prime - {a^\prime \over a} A_i \Big), \quad
       B_i = {1 \over a} \eta_{ijk} \partial^j A^k.
   \label{EB-A-Friedmann}
\eea

{\it Ignoring} the charge and current densities, from Eqs.\ (\ref{M-Friedmann-1})-(\ref{M-Friedmann-4}), using Eq.\ (\ref{EB-A-Friedmann}), we have
\bea
   & & \hskip -.8cm
       {1 \over a^4} ( a^2 B_i )^{\prime\prime}
       + \Big( {m^2 c^2 \over \hbar^2}
       - {\Delta \over a^2} \Big) B_i = 0,
   \label{B-eq-Friedmann} \\
   & & \hskip -.8cm
       {1 \over a^4} ( a^2 E_i )^{\prime\prime}
       + \Big( {m^2 c^2 \over \hbar^2}
       - {\Delta \over a^2} \Big) E_i
       = 2 {m^2 c^2 \over \hbar^2} {a^\prime \over a^2} A_i.
   \label{E-eq-Friedmann}
\eea
From Eqs.\ (\ref{A-0-Friedmann})-(\ref{Lorenz-Friedmann}), we have
\bea
   & & {1 \over a^3} \Big[ {1 \over a^2}
       ( a^3 A_0 )^\prime \Big]^\prime
       + \Big( {m^2 c^2 \over \hbar^2}
       - {\Delta \over a^2} \Big) A_0
       = 0,
   \label{A-0-eq} \\
   & & {1 \over a^2} ( a A_i )^{\prime\prime}
       - 2 {a^\prime \over a}
       {1 \over {m^2 c^2 \over \hbar^2} - {\Delta \over a^2}}
       {1 \over a^4}
       ( a A^j_{\;,j} )^\prime_{,i}
       + \Big( {m^2 c^2 \over \hbar^2}
       - {\Delta \over a^2} \Big) (a A_i)
       = 0.
   \label{A-i-eq}
\eea
The latter one can be decomposed into longitudinal and transverse parts as
\bea
   & & {1 \over a^2} ( a A^i_{\;,i} )^{\prime\prime}
       - 2 {a^\prime \over a} {\Delta \over a^2}
       {1 \over {m^2 c^2 \over \hbar^2}
       - {\Delta \over a^2}} {1 \over a^2}
       ( a A^i_{\;,i} )^\prime
       + \Big( {m^2 c^2 \over \hbar^2}
       - {\Delta \over a^2} \Big) (a A^i_{\;,i})
       = 0,
   \label{A-L-eq} \\
   & & \eta^{ijk} \partial_j \Big[ {1 \over a^2}
       ( a A_k )^{\prime\prime}
       + \Big( {m^2 c^2 \over \hbar^2}
       - {\Delta \over a^2} \Big) (a A_k) \Big] = 0.
   \label{A-T-eq}
\eea
Equations (\ref{A-L-eq}) and (\ref{A-T-eq}) for dark photons were studied in the context of inflation in \cite{Dimopoulos-2006}.

\subsection{Energy-momentum tensor}

In terms of EM fields, using the normal four-vector, Eq.\ (\ref{Tab-EM}) gives
\bea
   & & T_{00} = a^2 \Big\{ {1 \over 2}
       ( 1 + 2 \alpha - 2 \varphi ) ( E^2 + B^2 )
       - C^{ij} ( E_i E_j + B_i B_j )
       + {2 \over a} \eta_{ijk} E^i B^j \chi^k
   \nonumber \\
   & & \qquad
       + {m^2 c^2 \over \hbar^2} \Big[ {1 \over 2} A_0^2
       + {1 \over 2} ( 1 + 2 \alpha - 2 \varphi ) A^2
       - {1 \over a} \chi^i A_i A_0
       - C^{ij} A_i A_j \Big] \Big\},
   \nonumber \\
   & & T_{0i} = a^2 \Big\{
       (1 + \alpha - \varphi) \eta_{ijk} E^j B^k
       + 2 C_i^j \eta_{jk\ell} E^k B^\ell
       - {1 \over 2 a} \chi_i (E^2 + B^2)
       + {1 \over a} ( E_i E_j \chi^j + B_i B_j \chi^j )
   \nonumber \\
   & & \qquad
       + {m^2 c^2 \over \hbar^2} \Big[
       A_0 A_i + {1 \over 2 a} \chi_i ( A^2 - A_0^2 ) \Big] \Big\},
   \nonumber \\
   & & T_{ij} = a^2 \Big[ {1 \over 2} (E^2 + B^2)
       ( \delta_{ij} + 2 C_{ij} )
       - C^{k\ell} ( E_k E_\ell + B_k B_\ell ) \delta_{ij}
       - E_i E_j - B_i B_j \Big]
   \nonumber \\
   & & \qquad
       + a^2 {m^2 c^2 \over \hbar^2} \Big\{
       A_i A_j + \Big[  - {1 \over 2} A^2
       + {1 \over 2} ( 1 - 2 \alpha + 2 \varphi ) A_0^2
       + {1 \over a} \chi^k A_k A_0
       + C^{k\ell} A_k A_\ell \Big] \delta_{ij}
       + ( A_0^2 - A^2 ) C_{ij} \Big\},
\eea
where $T_{ab} = T^{\rm EM}_{ab}$.

For the homogeneous mode, we have $\bar A_0 = 0 = \bar B_i$, thus
\bea
   T_{00} = {a^2 \over 2} \Big( \bar E{}^2
       + {m^2 c^2 \over \hbar^2} \bar A{}^2 \Big), \quad
       T_{0i} = 0, \quad
       T_{ij} = a^2 \Big[ {1 \over 2} \bar E{}^2 \delta_{ij}
       - \bar E_i \bar E_j
       - {m^2 c^2 \over \hbar^2}
       \Big( {1 \over 2} \bar A{}^2 \delta_{ij}
       - \bar A_i \bar A_j \Big) \Big].
   \label{Tab-BG}
\eea
The above energy-momentum tensor is derived assuming homogeneous and isotropic background metric. This implies that for consistency the electromagnetic contribution should contribute negligibly to the background order energy-momentum tensor.

In the non-relativistic limit, with mass-dominated coherent oscillation stage of the field, $T_{ij} = 0$ and the energy-momentum content behaves as a dust in a homogeneous-isotropic background, see Eq.\ (\ref{fluid-NR}). This is known as the dark-photon dark matter \cite{Nelson-2011, Arias-2012, Ahmed-2020, Adshead-2021}. Previously, the case was derived as a non-relativistic limit of the perturbation theory. In the following we will derive the case as the Newtonian limit of the post-Newtonian approximation.

\end{widetext}
%
%
%
\section{Dark photon vector dark matter}
                                  \label{sec:Proca-DM}

In the non-relativistic limit with a coherent oscillation ansatz the dark Proca field behaves as a dark matter \cite{Nelson-2011, Arias-2012, Ahmed-2020, Adshead-2021}. Here, we derive the non-relativistic (Newtonian) limit of the Proca electrodynamics by using the post-Newtonian (PN) approximation. We consider the PN expansion of the metric as \cite{Chandrasekhar-1965, Hwang-Noh-Puetzfeld-2008}
\bea
   & & g_{00} = - \Big( 1 + {2 \over c^2} \Phi \Big), \quad
       g_{0i} = - a {1 \over c^3} P_i,
   \nonumber \\
   & & g_{ij} = a^2 \Big( 1 - {2 \over c^2} \Psi \Big)
       \delta_{ij},
   \label{metric-PN}
\eea
where $P_i$ and $\Psi$ terms are already 1PN order; $\Phi$ includes 1PN as well as 0PN order \cite{Chandrasekhar-1965}. In this subsection we use $x^0 = ct$, instead of $x^0 = \eta$ used in the previous section; in this notation we can use the PN quantities presented in Appendix B of \cite{Hwang-Noh-axion-2023}. We have $B_i$ etc. introduced as in Eq.\ (\ref{EB-cov-pert}) except for $\widetilde A_0 \equiv A_0$. Compared with the metric in perturbation theory in Eq.\ (\ref{metric-pert}), we have
\bea
   \alpha = {\Phi \over c^2}, \quad
       \varphi = - {\Psi \over c^2}, \quad
       \chi_i = a {P_i \over c^3},
\eea
and ignored tensor-type perturbation.

\subsection{Klein transformation}

For a scalar field, the Klein transformation transforms the scalar field equation of motion (Klein-Gordon equation) to a form of Schr\"odinger equation, and in the non-relativistic limit the Schr\"odinger equation is recovered \cite{Klein-1926}. Similarly, here we consider an {\it ansatz}
\bea
   A_a \equiv {\hbar \over \sqrt{2m}}
       ( {\cal A}_a e^{- imc^2t/\hbar}
       + {\cal A}^*_a e^{imc^2t/\hbar} ).
   \label{A-Psi}
\eea
We introduce ${\cal E}_i$ and ${\cal B}_i$ for $E_i$ and $B_i$, respectively, similarly as ${\cal A}_i$ for $A_i$. From Eq.\ (\ref{EB-A-cov}), to 1PN order, we have
\bea
   & & {\cal E}_i = {i m c \over \hbar} {\cal A}_i
       - {1 \over c} \Big(
       \dot {\cal A}_i + {\dot a \over a} {\cal A}_i
       + i {m \over \hbar} \Phi {\cal A}_i
       - {c \over a} {\cal A}_{0,i} \Big),
   \nonumber \\
   & & {\cal B}_i = {1 \over a} \eta_{ijk}
       \Big[ \Big( 1 + {1 \over c^2} \Psi \Big)
       \partial^j {\cal A}^k
       + i {a m \over \hbar c^2} P^j {\cal A}^k \Big].
\eea
We have ${\cal A}_0 \sim c^{-1} {\cal A}_i$, see Eq.\ (\ref{Lorenz-A-0PN}).

Now, we consider only to 0PN order. We have
\bea
   {\cal E}_i = {i m c \over \hbar} {\cal A}_i, \quad
       {\cal B}_i = {1 \over a} \eta_{ijk}
       \partial^j {\cal A}^k.
\eea
The Lorenz condition in Eq.\ (\ref{Lorenz-cov}) with Eq.\ (\ref{U}) gives
\bea
   - i {mc \over \hbar} {\cal A}_0
      = {1 \over a} {\cal A}^i_{\;,i}.
   \label{Lorenz-A-0PN}
\eea
Using this, the equation of motion in Eq.\ (\ref{EOM-A}) gives
\bea
   \Big[ i \Big( \partial_t
      + {3 \over 2} {\dot a \over a} \Big)
      + {\hbar \over 2 m} {\Delta \over a^2}
      - {m \over \hbar} \Phi
      \Big] {\cal A}_i = 0,
   \label{Schrodinger}
\eea
where we ignored charge and current densities. This is a set of Schr\"odinger equations for wave functions ${\cal A}_i$ in the context of expanding background with the gravitational potential $\Phi$. It shows that in non-relativistic limit the wave equations no longer depend on the polarization.

From Eq.\ (\ref{fluid-quantities}) using the fluid four-vector ($u_a$) and Eq. (\ref{Tab-EM}) using the normal four-vector ($n_a$), ignoring the oscillatory terms by taking average over time, fluid quantities become
\bea
   & & \varrho = m {\cal A}{}^i {\cal A}{}^*_i, \quad
       p = 0 = \pi_{ij},
   \nonumber \\
   & & \varrho v_i = {i \hbar \over 2 a} \big(
       {\cal A}_i {\cal A}^{j*}{}_{\hskip -.14cm ,j}
       - {\cal A}^*_i {\cal A}^j{}_{,j}
       + {\cal A}_{i,j} {\cal A}^{j*}
       - {\cal A}^*_{i,j} {\cal A}^j
   \nonumber \\
   & & \qquad
       + {\cal A}^*_{j,i} {\cal A}^j
       - {\cal A}_{j,i} {\cal A}^{j*} \big),
   \label{fluid-NR}
\eea
where $\mu = \varrho c^2$ and $v_i$ comes from imposing the energy-frame condition $q_a \equiv 0$; $v_i$ is introduced as $\widetilde u_i \equiv \gamma \widetilde v_i \equiv a \gamma v_i/c$ with the index of $v_i$ associated with $\delta_{ij}$, see Appendix B in \cite{Hwang-Noh-axion-2023} for derivation. Another way is to use Eq.\ (\ref{fluid-EM-cov}) using the fluid four-vector, and transform EM fields to the ones based on the normal four-vector using Eq.\ (45) in \cite{HN-2023-EM}. For a homogeneous mode, Eq.\ (\ref{Schrodinger}) gives ${\cal A}_i \propto a^{-3/2}$. We have $\varrho = m {\cal A}^i {\cal A}_i^* \propto a^{-3}$ and to the background order the field behaves as a dust fluid. We can show that the energy conservation equation is satisfied by the equation of motion and the momentum conservation equation is 1PN order.

\subsection{Madelung transformation}

For a scalar field, the Madelung transformation transforms the Schr\"odinger equation to the continuity and momentum conservation equations \cite{Madelung-1927}. Similarly, we take an {\it ansatz}
\bea
   {\cal A}_i \equiv \sqrt{\varrho^{(i)} \over m}
       e^{i m u^{(i)}/\hbar},
\eea
for each component of ${\cal A}_i$. The imaginary and real parts, respectively, of Eq.\ (\ref{Schrodinger}) give
\bea
   & & \hskip -.8cm
       \dot \varrho^{(i)} + 3 {\dot a \over a} \varrho^{(i)}
       + {1 \over a^2} ( \varrho^{(i)} u^{(i),k} )_{,k} = 0,
   \label{varrho-i-eq} \\
   & & \hskip -.8cm
       \dot u^{(i)}
       + {1 \over 2 a^2} u^{(i),k} u^{(i)}{}_{\hskip -.1cm ,k}
       + \Phi
       - {\hbar^2 \over 2 m^2 a^2}
       {\Delta \sqrt{\varrho^{(i)}} \over \sqrt{\varrho^{(i)}}}
       = 0.
   \label{u-i-eq}
\eea
Introducing ${\bf u}^{(i)} \equiv {1 \over a} \nabla u^{(i)}$, we have
\bea
   & & \hskip -.4cm
       \dot \varrho^{(i)} + 3 {\dot a \over a} \varrho^{(i)}
       + {1 \over a} \nabla \cdot
       ( \varrho^{(i)} {\bf u}^{(i)} ) = 0,
   \label{varrho-i-eq-2} \\
   & & \hskip -.4cm
       \dot {\bf u}^{(i)}
       + {\dot a \over a} {\bf u}^{(i)}
       + {1 \over a} {\bf u}^{(i)} \cdot \nabla {\bf u}^{(i)}
       + {1 \over a} \nabla \Phi
       - {\hbar^2 \over 2 m^2 a^3} \nabla
       {\Delta \sqrt{\varrho^{(i)}} \over \sqrt{\varrho^{(i)}}}
   \nonumber \\
   & & \qquad
       \hskip -.4cm
       = 0.
   \label{u-i-eq-2}
\eea
Previously, the above equations were derived as the non-relativistic limit of perturbation theory \cite{Adshead-2021}. Here, we derived these as the 0PN limit of the PN approximation.

For each $i$-component, Eq.\ (\ref{varrho-i-eq-2}) is the same as the mass conservation (or continuity) equation of Newtonian hydrodynamics in cosmological context. On the other hand, the momentum conservation equation of ordinary fluid hydrodynamics is
\bea
   \hskip -.3cm
   \dot {\bf u}
       + {\dot a \over a} {\bf u}
       + {1 \over a} {\bf u} \cdot \nabla {\bf u}
       + {1 \over a} \nabla \Phi
       + {1 \over a \varrho} ( \nabla_i p
       + \nabla_j \Pi^j_i ) = 0,
   \label{u-i-eq-3}
\eea
where $\Pi_{ij}$ is the anisotropic stress. The last term in Eq.\ (\ref{u-i-eq-2}) does not behave as a pressure gradient. Comparison of Eqs.\ (\ref{u-i-eq-2}) and (\ref{u-i-eq-3}) reveals that \cite{Takabayasi-1952}
\bea
   & & p^{(i)} = - {\hbar^2 \over 12 m^2 a^2}
       \varrho^{(i)} \Delta \ln{\varrho^{(i)}},
   \nonumber \\
   & & \Pi^{(i)}_{jk} = - {\hbar^2 \over 4 m^2 a^2}
       \varrho^{(i)} ( \nabla_j \nabla_k
        - {1 \over 3} \delta_{jk} \Delta ) \ln{\varrho^{(i)}}.
\eea
This shows that the last term in Eq.\ (\ref{u-i-eq-2}), being a mixture of pressure (isotropic stress) and anisotropic stress, can be interpreted as the {\it quantum stress} \cite{Takabayasi-1952}.

The naming `quantum' can be justified as the term involves $\hbar$. More importantly, the equation is derived from Eq.\ (\ref{Schrodinger}) which is apparently a set of Schr\"odinger equations. Although in a different context, David Bohm has shown that quantum properties of the Schr\"odinger equation, like the uncertainty principle, quantum tunneling, interference, and all the quantum aspects of Schr\"odinger equation are encoded in this quantum stress term \cite{Bohm-1952-I, Bohm-1952-II, Holland-1993}. The equivalence between the Schr\"odinger and hydrodynamic equations under the Madelung transformation, however, is challenged \cite{Wallstrom-1994}: the two formulations are different by lacking quantized circulation (vortex) surrounding regions of vanishing density (of the wave function) in the latter system, as is well known in superfluids and Bose-Einstein condensates \cite{Onsager-1949, Takabayasi-1952, Feynman-1955, Gross-1961, Pitaevskii-1961, Wallstrom-1994, Pethick-Smith-2002, Pitaevskii-Stringari-2003}.

We note that ${\bf u}^{(i)}$ {\it differs} from ${\bf v}$ based on the field's four-vector $u_a$ in Eq.\ (\ref{fluid-NR}). We have
\bea
   \varrho^{(i)} = m {\cal A}^i {\cal A}^*_i, \quad
       {\bf u}^{(i)} = {i \hbar \over 2 m a}
       \Big( { \nabla {\cal A}^*_i \over {\cal A}^*_i }
       - { \nabla {\cal A}_i \over {\cal A}_i } \Big),
\eea
where we do not sum over $i$-indices.

\subsection{Vector dark matter}

To perturbed order, we decompose $\varrho^{(i)} \rightarrow \varrho^{(i)} + \delta \varrho^{(i)} $ and $\delta \varrho^{(i)} \equiv \varrho^{(i)} \delta^{(i)}$. Equations (\ref{varrho-i-eq}) and (\ref{u-i-eq}) give,
\bea
   \dot \varrho^{(i)} + 3 {\dot a \over a} \varrho^{(i)} = 0,
\eea
to the background order, and
\bea
   \dot \delta^{(i)} + {\Delta \over a^2} u^{(i)} = 0, \quad
       \dot u^{(i)} + \Phi - {\hbar^2 \Delta \over 4 m^2 a^2} \delta^{(i)} = 0,
\eea
to the perturbed order. Combining these, we have
\bea
   \ddot \delta^{(i)} + 2 {\dot a \over a} \dot \delta^{(i)}
       + {\hbar^2 \Delta^2 \over 4 m^2 a^4} \delta^{(i)}
       - {\Delta \over a^2} \Phi = 0.
\eea
This can be combined with Poisson's equation which is a 0PN limit of Einstein's equation: considering only the field, we have
\bea
   {\Delta \over a^2} \Phi = 4 \pi G \delta \varrho
       = 4 \pi G \sum_k \varrho^{(k)} \delta^{(k)}.
\eea
To perturbed order the three density perturbation equations of $\delta^{(i)}$ are coupled through gravity.

For the same $\varrho^{(i)}$ and $\delta^{(i)}$ for each $i$, we have
\bea
   \ddot \delta^{(i)} + 2 {\dot a \over a} \dot \delta^{(i)}
       + \Big( {\hbar^2 \Delta^2 \over 4 m^2 a^4}
       - 4 \pi G \varrho \Big) \delta^{(i)} = 0,
\eea
where $\varrho = \sum_i \varrho^{(i)} = 3 \varrho^{(i)}$. Comparing the gravity term with the quantum stress term we have the Jeans scale. With $\Delta = - k^2$, we have
\bea
   \lambda_J = {2 \pi a \over k_J}
       = \pi \sqrt{ {\hbar \over m}
       {1 \over \sqrt{ \pi G \varrho } } }
       = { 66.4 {\rm kpc} \over
       \sqrt{ m_{22} h_{70} \sqrt{\Omega_0}} },
\eea
where $H \equiv 70 h_{70} {\rm km/sec/Mpc}$, $\Omega \equiv 8 \pi G \varrho/(3 H^2)$, $m_{22} \equiv m c^2/(10^{-22} {\rm eV})$, and $\Omega_0$ and $\varrho_0$ indicate the present values; Compton wavelength is $\lambda_C \equiv h/(mc) = 0.40{\rm pc}/m_{22}$. These are the same as the scalar dark matter \cite{Hwang-Noh-axion-2022}. Therefore, depending on the mass scale the dark Proca field behaves as the cold or fuzzy dark matter just like its scalar counter part of the axion or axion-like particles \cite{Hu-2000, Marsh-2016, Niemeyer-2020, Ferreira-2021, Hui-2021}.

%
%
%
\section{Discussion}
                                  \label{sec:Discussion}

We studied massive-photon electrodynamics and MHD in the context of general curved spacetime and in perturbed cosmological spacetime.

Some new aspects are the following. (i) The massive photon electrodynamics and MHD are presented in general curved spacetime of Einstein's gravity using the covariant and the ADM formulations. (ii) Maxwell equations with Proca mass term in perturbed Friedmann cosmology are presented without fixing the temporal gauge condition. (iii) The gauge issue of the EM fields and vector potential in linearly perturbed cosmology is clarified. (iv) Wave equations for the EM fields and vector potential are derived assuming derivatives of gravity are negligible compared with derivatives of fields. (v) The vector dark matter nature of dark Proca field in the non-relativistic limit under the Klein transformation is studied using the post-Newtonian approximation.

The Maxwell's and MHD equations in the covariant and the ADM formulations are applicable to general curved spacetime in Einstein's gravity. Relativistic astrophysics involving magnetic field is an active field with many textbooks on the subject often based on the ADM formulation \cite{Thorne-1986, Wilson-Mathews-2003, Baumgarte-Shapiro-2010, Beskin-2010, Gourgoulhon-2012, Shibata-2015, Baumgarte-Shapiro-2021}. We have added the Proca mass term to the relativistic MHD in the ADM formulation. Even without the Proca mass term, our ADM equations for Maxwell's equations in terms of the EM fields and the field potential may have new aspects waiting to be investigated.

Due to the anisotropy caused by the vector nature of the Maxwell field, the gravity of the vector field cannot be accommodated within the spatially homogeneous and isotropic Friedmann background. One important exception is the case of non-relativistic limit under the Klein transformation where a dark Proca field behaves as vector dark matter in the Friedmann world model. We studied the case using the post-Newtonian approximation. In this limit, to the 0PN order, the polarization dependence disappears and depending on the mass of the field it can behave as the cold or fuzzy dark matter. The 1PN order will include the first relativistic correction; for the axion dark matter case, see \cite{Hwang-Noh-axion-2023}.

The currently reliable experimental limit on the photon mass is negligibly small in the laboratory standard. However, considering the speculative status of the photon-mass bound based on astrophysical magnetic fields beyond the Solar System scale, with the photon Compton wavelength lower bound merely around 1.3AU, roles of the photon-mass in the larger scale astrophysical processes may deserve more attention. It might be reasonable to consider the effect of photon mass in all the electromagnetic and MHD processes beyond the current limit of photon Compton wavelength. The electrodynamic and MHD equations in general curved spacetime and perturbed cosmology contexts presented here might be useful for future applications.

%
%
%
\vskip .5cm
\centerline{\bf Acknowledgments}

We wish to thank Professors Kiwoon Choi, Donghui Jeong, Taehyun Jung and Pyungwon Ko for sharing their insights on the subject. We thank two anonymous referees for useful suggestions to improve presentation and Professor Alessandro D.A.M. Spallicci for guiding us to historical literature and providing useful references. H.N.\ was supported by the National Research Foundation (NRF) of Korea funded by the Korean Government (No.RS-2024-00333721 and No.2021R1F1A1045515). J.H.\ was supported by IBS under the project code, IBS-R018-D1.

%
%


\end{document}